\documentstyle[12pt]{article}
\def\ba{\begin{eqnarray}}
\def\ea{\end{eqnarray}}

\def\lb{\label}
\def\be{\begin{equation}}
\def\ee{\end{equation}}

\def\C{\mathbb C}
\begin{document}

\title{New $G_2$ holonomy metrics, D6 branes with inherent $U(1)\times U(1)$
isometry and $\gamma$-deformations}
\author{O.P.Santillan \thanks{osvaldo@theor.jinr.ru and firenzecita@hotmail.com}}
\date{}
\maketitle

\begin{abstract}
It is found the most general local form of the 11-dimensional
supergravity backgrounds which, by reduction along one isometry,
give rise to IIA supergravity solutions with a RR field and a non
trivial dilaton, and for which the condition $F^{(1,1)}=0$ holds.
This condition is stronger than the usual condition
$F^{ab}J_{ab}=0$, required by supersymmetry. It is shown that these
D6 wrapped backgrounds arise from the direct sum of the flat
Minkowski metric with certain $G_2$ holonomy metrics admitting an
$U(1)$ action, with a local form found by Apostolov and Salamon.
Indeed, the strong supersymmetry condition is equivalent to the
statement that there is a new isometry on the $G_2$ manifold, which
commutes with the old one; therefore these metrics are inherently
toric. An example that is asymptotically Calabi-Yau is presented.
There are found another $G_2$ metrics which give rise to half-flat
$SU(3)$ structures. All this examples possess an $U(1)\times
U(1)\times U(1)$ isometry subgroup. Supergravity solutions without
fluxes corresponding to these $G_2$ metrics are constructed. The
presence of a $T^3$ subgroup of isometries permits to apply the
$\gamma$-deformation technique in order to generate new supergravity
solutions with fluxes.

\end{abstract}

\tableofcontents

\section{Introduction}

   The AdS/CFT correspondence is a powerful tool in order to study strongly coupled
regimes in gauge theories \cite{Juanma}. It relates field theories
with gravitational theories satisfying particular boundary
conditions. The original statement of the correspondence was that
${\cal N}=4$ super-Yang Mills theory is dual to type IIB strings in
$AdS^5\times S^5$. This pioneer conjecture was developed further in
\cite{Klebanov} and even it was generalized to non conformal field
theories \cite{Yankie}.

   In the particular case of ${\cal N}=4$ super-Yang Mills
there exist a three parameter deformations of its superpotential
that preserves ${\cal N}=1$ supersymmetry, these deformations are
called $\beta$ deformations \cite{Leigh}. The original
superpotential of the theory is transformed in the following way \be
\lb{deaf} Tr( \Phi_1 \Phi_2 \Phi_3 -\Phi_1 \Phi_3 \Phi_2 ) \to h \,
Tr( e^{ i  \pi \beta} \Phi_1\Phi_2 \Phi_3 - e^{- i  \pi \beta}
 \Phi_1\Phi_3 \Phi_2  ) + h' Tr( \Phi_1^3 +\Phi_2^3 +\Phi_3^3 ),
\ee
being $h, h' , \beta$ complex parameters, satisfying one condition
by conformal invariance. One can select $h'=0$.
Besides the $U(1)_R$ symmetry,  there is
a  $U(1) \times U(1)$ global symmetry generated by
$$
U(1)_1:\;\;(\Phi_1,\Phi_2 ,\Phi_3 ) \to (\Phi_1,e^{i\varphi_1}\Phi_2
,e^{-i \varphi_1} \Phi_3 ),
$$
\be\lb{mmm} U(1)_2:\;\;(\Phi_1,\Phi_2 ,\Phi_3 ) \to (e^{-i
\varphi_2} \Phi_1,e^{i\varphi_2}\Phi_2 , \Phi_3 ), \ee which leaves
the superpotential and the supercharges invariant. Therefore there
is a two dimensional manifold of ${\cal N}=1$ CFT with a torus
symmetry. The physics contained in the deformed model is periodic in
the variable $\beta$, which is parameterized as \be \lb{gams} \beta
= \gamma - \tau_s \sigma \ee where $\gamma$ and $\sigma$ are real
variables with period one. The variable $\beta$ can be considered
living on a torus with complex structure $\tau_s$, where $\tau_s $
is related to gauge coupling and theta parameter of the field theory
\cite{Dorey}. The theory has an $SL(2,Z)$ duality group, in which
$\beta$ transforms as a modular form \be \lb{sdualo} \tau_s \to { a
\tau_s + b \over c \tau_s + d} ~,~~~~~~~~~\beta \to { \beta
 \over c\tau_s + d} ~,~~~~~~~~~~ \left( \begin{array}{cc}  a & b \\
  c & d \end{array} \right) ~\in  SL(2,Z).
\ee If $\beta$ is chosen to be real then $\sigma=0$. We will focus
mainly in the case $\sigma=0$, the corresponding deformations are
called $\gamma$ deformations.

   An interesting problem is to know how these deformations
acts on the gravity dual, which has an $U(1)\times U(1)$ subgroup
which is realized as an isometry. The answer is that the
$\gamma$-deformation of ${\cal N}=4$ super Yang-Mills induce in the
gravity dual the simple transformation \be \lb{tau} \tau\equiv B + i
\sqrt{g} \longrightarrow \tau \longrightarrow \tau' =\frac { \tau}{
1 + \gamma \tau}, \ee where $\sqrt{g}$ is the volume of the two
torus \cite{Lunin}. The transformation (\ref{tau}) is in fact, a
well known solution generating technique \cite{Horowitz}. It is well
known that when a closed string theory is compactified on a two
torus the resulting eight dimensional theory has an exact $SL(2,Z)
\times SL(2,Z)$ symmetry, which acts on the complex structure of the
torus and on certain parameter \be\lb{metrr} \tau = B_{12} + i
\sqrt{g}, \ee where $\sqrt{g}$ is the volume of the two torus in
string metric. At supergravity level there is an enlarged $SL(2,R)
\times SL(2,R)$ symmetry, which is not a symmetry of the full string
theory. These  $SL(2,R)$ symmetries can be used as solution
generating transformation, at supergravity level.

   As the reader can check, the transformations (\ref{tau})
are not the full $SL(2,R)$ transformations. Indeed (\ref{tau}) is
the subgroup of $SL(2,R)$ for which $\tau\to 0$ implies that
$\tau'\to 0$. It can be directly checked from (\ref{tau}) that
$\tau_\gamma = \tau + o(\tau^2) $ for small $\tau$. Transformations
with these properties are the only possible ones mapping a ten
dimensional geometry which is non singular to a new one also without
singularities. The reason is that the only points where a
singularity can be introduced by performing an $SL(2,R)$
transformation is where the two torus shrinks to zero size. This
shrink happens when $\tau' \to 0$ but for $\gamma$ transformations,
this implies that $\tau\to 0$. Therefore, if the original metric was
non-singular, then the deformed metric is also non singular
\cite{Lunin}. The transformation (\ref{tau}) is the result of doing
a T-duality on one circle, a change of coordinates, followed by
another T-duality. This is another reason for which it can be
interpreted as a solution generating technique \cite{Horowitz}.

    In the present work new $G_2$ holonomy metrics will be constructed and
the deformation procedure will be used to generate new supergravity
backgrounds. The local form of such $G_2$ metrics is \cite{Apostol}
\be\lb{pinpin} g_{7}=\frac{(d\chi+A_2)^2}{\theta^2}+\theta\;[\;u \;
d\theta^2+\frac{(d\upsilon+A_1)^2}{u}+g_{4}(\theta)\;]. \ee All the
quantities appearing in (\ref{pinpin}) are independent on the
coordinates $\upsilon$ and $\chi$, and the vector fields
$\partial_{\chi}$ and $\partial_{\upsilon}$ are Killing and
commuting. Therefore by construction these metrics are toric. The
metric $g_{4}(\theta)$ is a 4-metric at each level of constant
$\theta$-surfaces. If $g_{4}(\theta)$ admit more isometries, then
the full isometry group of (\ref{pinpin}) can be enlarged sometimes.
In certain situations examples with $T^3$ or larger isometry groups
can be generated. The undefined quantities appearing in the
expression (\ref{pinpin}) are not arbitrary, the condition of
holonomy to be included in $G_2$ gives an system of evolution
equations relating all them \cite{Apostol}.

    There are different reasons for taking these 7-metrics
as starting point. The general form (\ref{pinpin}) encodes known
examples as particular subcases. For instance, the homothetic $G_2$
metrics, arising from an $SO(5)$ invariant $G_2$ metric by
contraction of the isometry group, with local form
$$
g_7= \frac{(d\upsilon-x dz + y d\varsigma)^2}{\theta^2}+
\frac{(d\chi-y dz-x d\varsigma)^2}{\theta^2}
+\theta^4\;d\theta^2 +\theta^2\;(\;dx^2+dy^2+dz^2+d\varsigma^2\;),
$$
found in \cite{Stele}, is an special subcase of (\ref{pinpin}). More
general examples are also included. There are known some explicit
examples of $G_2$ holonomy metrics \cite{Bryant}-\cite{Lu}, and one
of the purpose of the present work is to enlarge the list.

     There is another reason to consider
the metrics (\ref{pinpin}). As is known, the problem of finding the
conditions for preserving supersymmetry in type IIA string theory in
the presence of a RR vector field and a nontrivial dilaton can be
derived from reduction of the supersymmetry conditions of
eleven-dimensional supergravity along certain isometry. If a further
constraint $F^{(1,1)}=0$ is imposed, which is stronger than the
usual condition $F^{ab}J_{ab}=0$, required by supersymmetry, then
the ten-dimensional string frame metric in the presence of the
D6-branes contains an internal 6-space that is Kahler. As it was
shown in \cite{Kaste} such kind of geometries are characterized by
certain holomorphic monopole equation. By use of the results given
in \cite{Apostol} we will prove that indeed the metrics
(\ref{pinpin}) parameterize the solution of the holomorphic monopole
equation and the constraint $F^{(1,1)}=0$. It will be shown that
both conditions implies that a second isometry is inherently defined
on both the 7-dimensional $G_2$ manifold and the 6-dimensional
Kahler manifold. Therefore the $G_2$ metrics characterized by these
conditions are inherently \emph{toric}, and in some cases they admit
a larger group of isometries like $U(1)^3$ or others. One can
immediately construct backgrounds that are suitable to apply the
$\gamma$ solution generating technique. We are focused in the
gravity part of the conjectured duality, mainly because we do not
know yet the quantum field theory dual of our backgrounds.

   In the first part of the present work it will be shown the reason for
which these are the metrics solving the condition $F^{(1,1)}=0$ and
the holomorphic monopole equation. Explicit examples will be found,
some of them are known and other are new. In the second part the
$\gamma$-deformed backgrounds corresponding to these $G_2$ holonomy
metrics are presented.

\section{Toric D6 brane backgrounds with strong supersymmetry conditions}

\subsection{Wrapped IIA solutions with Kahler internal geometry}

  It is an standard fact that if the holonomy of a seven manifold
$Y$ is included in $G_2$, then from the very beginning there exist
at least one globally defined covariantly constant spinor $\eta$
over $Y$, that is, an spinor satisfying $D\eta=0$ over $Y$, being
$D$ standard covariant derivative in the representation of the
field. This fact implies that $Y$ is Ricci-flat. It also implies
that the curvature two-form $R_{ab}$ of $Y$ is octonion self-dual,
that is
$$
R_{ab}=\frac{c_{abcd}}{2}R_{cd},
$$
being $c_{abcd}$ the dual octonion multiplication constants. The
self-duality of the curvature implies the existence of a frame for
which the spin connection $\omega_{ab}$ is also self-dual. All these
conditions are equivalent to the existence of a $G_2$ invariant
closed and co-closed three form $\Phi$ \cite{Bryant}.

    Along the present section, a generic solution of the
eleven dimensional supergravity in which the fermions and the four
form $F$ are zero is considered. Such solution is of the form
\be\lb{kasto} g_{11}=g_{(3,1)}+g_{7}. \ee The metric $g_7$ is
defined on a seven manifold $Y$, and we assume that is non compact,
with holonomy included in $G_2$ and that there is at least one
$U(1)$ isometry. Without loosing generality, the Killing vector
field can be written as $\partial_{\chi}$, being $\chi$ certain
coordinate. Then the full supergravity background takes the usual
IIA form \be\lb{mert} g_{11}= e ^{-2\alpha \phi} g_{10} + e ^{2\beta
\phi}(d\chi+A)^2 = g_{(3,1)}+ g_7. \ee The metric $g_{10}$ is the
physical metric IIA in ten dimensions obtained by reduction along
the $U(1)$ isometry; the parameters $\alpha$ and $\beta$ determine
the frame, with values $(\alpha,\beta)=(1/3,2/3)$ for the string
frame. The $G_2$ metric can be decomposed as \be\lb{meri} g_{7} = e
^{-2\alpha\phi}g_{6} + e ^{2\beta \phi}(d\chi+A)^2, \ee and the
expression for $g_{10}$ is \be\lb{diuha} g_{10}=e ^{2\alpha \phi}
g_{(3,1)}+ g_{6}. \ee The metric $g_6$ is a six dimensional metric
defined over a manifold $N$. The pair $(N,g_6)$ the Riemannian
quotient of $(Y, g_7)$ by the $U(1)$ isometry, so that $N$ is a
6-dimensional manifold formed from the orbits of the Killing vector
field $V=\partial_{\chi}$. The IIA reduction gives the background
\be\lb{redu} g_{10}=e ^{2\alpha \phi} g_{(3,1)}+ g_{6}, \qquad F=dA
\ee being $F$ the RR two form. The coordinate $\phi$ is interpreted
as the dilaton field. As is was shown in \cite{Kaste}, the
supersymmetry condition $D\eta=0$ implies the following system for
the six dimensional manifold \be\lb{devil}
F^{ab}\overline{J}_{ab}=0,\qquad d(e^{-2\beta\phi})=-
\ast(e^{-\alpha\phi}\psi_3 \wedge F). \ee The second (\ref{devil})
is known as an holomorphic monopole equation in the terminology of
the reference \cite{Kaste}. The following part is intended to find
an explicit form of the $G_2$ holonomy metric (\ref{meri}) (and an
explicit form of the background $g_{10}$) under the assumption that
$g_6$ is Kahler. The Kahler condition implies that $F^{(1,1)}=0$ for
the six dimensional manifold, which is a requirement stronger that
the first (\ref{devil}).

\subsubsection{The IIA reduction and holomorphic monopole equations}

   If the internal manifold
has holonomy in $G_2$, then the four dimensional theory obtained by
dimensional reduction over the $G_2$ background will have ${\cal
N}=1$ supersymmetry. As is sketched above, the $G_2$ holonomy
condition implies the existence of a closed and co-closed form
$\Phi$ defined over our seven manifold $Y$. It always exist a
seven-bein $e^a$ for which the metric $g_7$ is written in diagonal
form $g=\delta_{ab}e^a\otimes e^b$, and for which the three form
$\Phi$ is expressed as \be\lb{lafo} \Phi=\frac{1}{3!} c_{abc} e^a
\wedge e^b \wedge e^c, \ee being $c_{abc}$ the octonion
multiplication constants. The fact that (\ref{lafo}) is $G_2$
invariant is a consequence that $G_2$ is the automorphism group of
the octonions. The holonomy of $Y$ will be included in $G_2$ if and
only if \cite{Bryant} \be\lb{holon} d\Phi=d\ast\Phi=0. \ee We will
suppose that the Killing vector $V$ not only preserve $g_7$ but also
$\Phi$. It means that the three form $\Phi$ is decomposed as
\be\lb{dico} \Phi= e ^{-3\alpha\phi} \psi_3 + e ^{-2\alpha \phi}
\overline{J} \wedge e ^z, \ee being $e^z=d\chi+A$ the co-tangent
vector associated to the $\chi$ direction and $\psi_3$ certain
3-form, and the factors associated to the dilaton $\phi$ in
(\ref{dico}) have been introduced by convenience. Then the two form
$\overline{J} =i_V \Phi$ satisfy \be\lb{horizonte} i_V
\overline{J}=0, \ee where $i_{V}$ denote the contraction with the
vector field $V$. A two form satisfying (\ref{horizonte}) is called
horizontal and as a consequence of horizontality and that $V$, by
assumption, preserve the $G_2$ structure, it is obtained that
$$
d\overline{J}=d(i_V \Phi)={\cal
L}_{V} \Phi=0,
$$
\be\lb{solon} {\cal L}_{V}\overline{J}=d(i_{V} \overline{J})=0. \ee
The first of conditions (\ref{solon}) implies that $\overline{J}$ is
closed, and the second that $\overline{J}$ it is preserved by $V$.
It also hold the decomposition \be\lb{dico2} \ast \Phi= e
^{-\alpha\phi}\psi'_3 \wedge e^z+ e ^{-2\beta\phi} \overline{J}
\wedge \overline{J}, \ee where the 3-form $\psi'_3$ is defined by
\be\lb{dim} e ^{-\alpha\phi}\psi'_3=i_{V}(\ast\Phi). \ee The form $e
^{-\alpha\phi}\psi'_3$ is also closed and preserved by $V$, the
proof has similarities with the one for $\overline{J}$. The
antisymmetric tensor $J$ defined through the relation $g_6(J \cdot,
\cdot)=\overline{J}(\cdot,\cdot)$ is an almost complex structure,
that is, it satisfies $J\cdot J=-I$. From (\ref{solon}) it is seen
that the vector $V$ preserve $\overline{J}$, such isometries are
called holomorphic. But a Killing vector preserving $\overline{J}$
always preserve $J$, such isometries are known as hamiltonian
isometries. Therefore $V$ is a Killing, hamiltonian and holomorphic
vector field.

   Due to the fact that
$\Phi\wedge \ast\Phi\sim \omega_{7}$, the three forms $\psi_3$ and
$\psi'_3$ together with $\overline{J}$ should satisfy the following
compatibility conditions \be\lb{compat} \psi_3\wedge
\psi'_3=\frac{2}{3}\overline{J}\wedge \overline{J}\wedge
\overline{J}=4 \omega_{6}, \ee
$$
\psi_3\wedge \overline{J}=\psi'_3\wedge \overline{J}=0,
$$
and therefore they conform an $SU(3)$ structure. Here $\omega_6$ and
$\omega_{7}$ are the volume forms on $Y$ and $N$ respectively. The
theory of $G_2$ structures gives the relations \cite{Chiozzi}
\be\lb{don}
\psi'_3(\cdot,\cdot,\cdot)=-\psi'_3(J\cdot,J\cdot,\cdot),\qquad
\psi_3(\cdot,\cdot,\cdot)=\psi'_3(J\cdot,\cdot,\cdot) \ee From the
first (\ref{don}) it is deduced that $\psi'_3$ has type
$(0,3)+(3,0)$ with respect to $J$, and from the second it follows
that the complex three form $\psi=\psi_3+i\psi'_3$ is of $(0,3)$
type with respect to $J$.

    In \cite{Kaste}-\cite{Stele} there were worked out the consequences of the
$G_2$ conditions (\ref{holon}) for a manifold with a Killing vector
$V$ preserving the $G_2$ structure, as in our case. It was found
that one can reduce to six dimensions these equations and divide the
resulting equations into pieces containing or not $e^z$. The system
that is finally obtained is the following
$$
d(e^{-3\alpha\phi}\psi_3) + e ^{(\beta-2\alpha)\phi} \overline{J}\wedge F=0,
\qquad d( e^{(\beta-2\alpha)\phi} \overline{J})=0,
$$
\be\lb{choga}
d(e^{-4\alpha\phi}(\ast \overline{J})) - e^{(\beta-3\alpha)\phi} (\ast\psi_3) \wedge F=0\ ,
\qquad d( e^{(\beta-3\alpha)\phi} (\ast \psi_3))=0.
\ee
The Hodge operation $\ast$ is referred the physical metric
$g_6$ and we have defined the "strength" 2-form $F=dA$.

  Using the relation $\beta=2\alpha$ of the ten-dimensional string frame
it is obtained from the second equation (\ref{choga}) that
$d\overline{J}=0$. Therefore $J$, $\overline{J}$ and $g_6$
constitute an almost Kahler structure. As is well known, this
structure will be Kahler if and only if the complex structure $J$ is
integrable, that is, if its Niejenhuis tensor
$$
N^{J}(X,Y)=[X,Y]+J[X,JY]+J[JX,Y]-[JX, JY]
$$
vanish identically. Let us assume that the Kahler condition hold and
derive its consequences. If this is so, then the manifold is complex
and it always exist a system of complex coordinates ($z^a$,
$\overline{z}^{\overline{a}}$) for which the tensors $\overline{J}$,
$e^{-\alpha\phi} \psi_3$ and $e^{-\alpha\phi} \psi_3'$ take the
simple form
$$
\overline{J}=\frac{i}{2} dz^a \wedge d\overline{z}^{\overline{a}},
$$
\be\lb{jon}
e^{-\alpha\phi}\psi_3= \frac{1}{2}( \frac{1}{3!} \epsilon_{abc}dz^{a}\wedge dz^b\wedge dz^{c} +
 \frac{1}{3!} \epsilon_{\overline{a}\overline{b}\overline{c}}d\overline{z}^{\overline{a}}
\wedge d\overline{z}^{\overline{b}}\wedge d\overline{z}^{\overline{c}}),
\ee
$$
\ast e^{-\alpha\phi}\psi_3= \frac{i}{2} ( \frac{1}{3!} \epsilon_{abc}dz^{a}\wedge dz^b \wedge dz^{c} -
 \frac{1}{3!} \epsilon_{\overline{a}\overline{b}\overline{c}}d\overline{z}^{\overline{a}}
\wedge d\overline{z}^{\overline{b}}\wedge d\overline{z}^{\overline{c}}),
$$
being $dz^a$ and $d\overline{z}^{\overline{a}}$ frames on the
holomorphic and anti-holomorphic cotangent bundle, respectively. The
integrability condition implies that these forms are all closed. In
this system of coordinates it is also simpler to see that the
$(2,2)$ part of the first equation (\ref{choga}) implies that
$F^{(1,1)}\wedge \overline{J}=0$, from where we obtain the well
known result that a Kahler reduction implies that \cite{Kaste}
\be\lb{strong} F^{(1,1)}=0, \ee a condition stronger than the usual
one $F^{ab}J_{ab}=0$, required by supersymmetry. We will to
(\ref{strong}) as an strong supersymmetry preserving condition. If
this condition is not satisfied, then the internal manifold $N$ will
be an $SU(3)$ torsion manifold in general \cite{Chiozzi}.

     The $(3,1)$ and $(1,3)$ components of the first equation
(\ref{choga}), together with the fourth equation, give the
generalized monopole equations \be\lb{monop} d(e^{-2\beta\phi})=-
\ast(e^{-\alpha\phi}\psi_3 \wedge F), \ee which in the basis defined
by $dz^a$ and $d\overline{z}^{\overline{a}}$ takes the form
\be\lb{toco}
\partial(e^{-2\beta\phi})=-\ast(e^{-\alpha\phi}\psi_3^{(3,0)}
\wedge F^{(2,0)}). \ee The third equation (\ref{choga}) can be
rewritten in the form \be\lb{monio} d(\ast\overline{J})-4\alpha
d\phi \wedge \ast \overline{J}-F\wedge \ast \psi_3=0, \ee but $\ast
\overline{J}=\frac{1}{2}\overline{J}\wedge \overline{J}$ and this,
together with the closure of $J$ implies that $d(\ast
\overline{J})=0$. The sum of the last two terms of (\ref{monio})
results again into another form of the monopole equations
(\ref{monop}). Equation (\ref{choga}) also implies that the internal
six manifold admits a gauge covariantly constant spinor playing the
role of supersymmetry generator \cite{Kaste}. The system
(\ref{choga}) also characterize the sub-manifold $M$ which the $N$
supersymmetric D6-branes wrap. Being a magnetic source of charge $N$
for the two form $F$, one has $dF = N \delta_M$, where $\delta_M$ is
the Poincare dual three-form of the cycle $M$. One can deduce from
(\ref{choga}) that \be\lb{fion} J\wedge \delta_M =0, \qquad e
^{-\alpha\phi} (\ast \psi_3) \wedge \delta_M =0, \ee and the first
of these equations implies that $M$ is a Lagrangian cycle in the
internal Kahler manifold.
\\

\textit{A new isometry from the strong supersymmetry condition}
\\

 The next task is to work out in detail the consequences of all
the equations presented above. It is convenient, in order to
simplify the following formulas and in order to compare them with
those appearing in the mathematical literature, to define a new
field (or a new coordinate) \be\lb{dili}
\theta=e^{-2\alpha\phi}=e^{-\beta\phi}, \qquad
\phi=-\frac{1}{2\alpha}\log(\theta), \ee which is entirely defined
in terms of the dilaton. The monopole system (\ref{monop}) can be
rewritten as
 \be\lb{monio} d(\ast\overline{J})-2
\log(\theta)\wedge \ast \overline{J}-F\wedge \ast \psi_3=0. \ee
Equation (\ref{choga}) is also reexpressed as  \be\lb{cosn}
\overline{J}\wedge dA+\theta^{1/2}d\theta \wedge \psi_3=0. \ee Now
comes the crucial point. Let us define a vector field $U$ by
\be\lb{hamo} i_{U}\overline{J}=-d\theta. \ee From (\ref{hamo}) it is
concluded that
$$
{\cal L}_{U} \overline{J}=i_{U} d\overline{J}+ d(i_{U}
\overline{J})=d(i_{U} \overline{J})=-d(d\theta)=0,
$$
being $i_{U}$ the contraction with the field $U$. Therefore ${\cal
L}_{U} \overline{J}=0$ and $U$ is an holomorphic vector field. We
will prove now the \emph{crucial} formula \be\lb{ifo}
dA=-\theta^{1/2}i_{U}\psi_3, \ee which holds as a direct consequence
of the (1,1) part of (\ref{cosn}). A good consistency check of
(\ref{ifo}) comes from realizing that combining (\ref{hamo}) and
(\ref{ifo}) with (\ref{cosn}) gives
$$
\overline{J}\wedge dA+\theta^{1/2}d\theta \wedge
\psi_3=-\overline{J}\wedge \theta^{1/2}i_{U}\psi_3
-\theta^{1/2}i_{U}\overline{J} \wedge
\psi_3=-\theta^{1/2}i_{U}(\overline{J}\wedge \psi_3)=0,
$$
due to the second $SU(3)$ condition (\ref{compat}). Therefore
(\ref{ifo}) implies (\ref{cosn}). We need to prove the converse,
that is, that (\ref{cosn}) implies (\ref{ifo}). Let us suppose that
$U$ is unitary (otherwise we will multiply it by its norm). An
$SU(3)$ basis $(e_i, Je_i)$ can be defined, in which $e_1=U$ and
$Je_1=JU$. Then the complex basis $(z_i, \overline{z}_i)$ with
$z_i=e_i-iJe_i$ can be constructed. In particular $z_1=U-iJU$,
$\overline{z}_i=z_i^*$ and $U=z_1+\overline{z}_1$. The dual basis is
denoted here with upper indices as $(z^i, \overline{z}^i)$. The
expression of $\overline{J}$ and $\theta^{1/2}\psi_3$ in this basis
is \be\lb{mur} \overline{J}=z^1\wedge \overline{z}^1 + z^2\wedge
\overline{z}^2 + z^3\wedge \overline{z}^3,\qquad \theta^{1/2}\psi_3=
-\Im(z^1\wedge z^2\wedge z^3), \ee where $\Im$ denotes the imaginary
part. The relation (\ref{hamo}) implies that
$$
d\theta=-i_U \overline{J}=-i_{(z_1+\overline{z}_1)}\overline{J}=
-i_{z_1}\overline{J}-i_{\overline{z}_1}\overline{J},
$$
and from the last expression together with the first (\ref{mur}) it
is found that
$$
d\theta(\overline{z}_j)=-\overline{J}(z_1, \overline{z}_j), \qquad
d\theta(z_j)=-\overline{J}(\overline{z}_1, z_j),
$$
and this implies that
$$
d\theta(z_2)=d\theta(\overline{z}_2)=d\theta(z_3)=d\theta(\overline{z}_3)=0,
$$
so that, evaluating the equality (\ref{cosn}) at the vectors
$(\overline{z}_j, z_j, z_i, z_k)$, $(z_j, \overline{z}_j,
\overline{z}_i, \overline{z}_k)$, $(z_j, \overline{z}_j,
\overline{z}_i, z_k)$ and $(z_j, \overline{z}_j, z_i,
\overline{z}_k)$ and using the second (\ref{mur}) gives
\be\lb{chacha} dA (z_i, z_k) = -\theta^{-1/2}\delta_{1j} \psi_3(z_j,
z_i, z_k), \qquad dA (\overline{z}_i, \overline{z}_k) =
-\theta^{-1/2}\delta_{1j} \psi_3(\overline{z}_j, \overline{z}_i,
\overline{z}_k) \ee
$$
dA (\overline{z}_i, z_k) = -\theta^{-1/2}\delta_{1j}
\psi_3(\overline{z}_j, \overline{z}_i, z_k)=0, \qquad dA (z_i,
\overline{z}_k) = -\theta^{-1/2}\delta_{1j} \psi_3(z_j, z_i,
\overline{z}_k)=0.
$$
The last two expressions shows that $F=dA$ is of type $(2,0) +
(0,2)$, which is in agreement with $F^{(1,1)}=0$. We also have that
the components
$$
\psi_3(\overline{z}_j, z_i, z_k)=\psi_3(z_j, \overline{z}_i,
\overline{z}_k)=\psi_3(\overline{z}_j, \overline{z}_i,
z_k)=\psi_3(\overline{z}_j, z_i, \overline{z}_k)=0,
$$
and so they can be conveniently added to (\ref{chacha}) to give
$$
dA (z_i, z_k) = -\theta^{-1/2}\delta_{1j} \psi_3(z_j, z_i,
z_k)-\theta^{-1/2}\delta_{1j} \psi_3(\overline{z}_j, z_i, z_k),
$$
$$
dA (\overline{z}_i, \overline{z}_k) = -\theta^{-1/2}\delta_{1j}
\psi_3(\overline{z}_j, \overline{z}_i,
\overline{z}_k)-\theta^{-1/2}\delta_{1j} \psi_3(z_j, \overline{z}_i,
\overline{z}_k),
$$
and also
$$
dA (\overline{z}_i, z_k) = -\theta^{-1/2}\delta_{1j}
\psi_3(\overline{z}_j, \overline{z}_i, z_k)-\theta^{-1/2}\delta_{1j}
\psi_3(z_j, \overline{z}_i, z_k)=0,
$$
$$
dA (z_i, \overline{z}_k) = -\theta^{-1/2}\delta_{1j} \psi_3(z_j,
z_i, \overline{z}_k)-\theta^{-1/2}\delta_{1j} \psi_3(\overline{z}_j,
z_i, \overline{z}_k)=0,
$$
from where it clearly follows (\ref{ifo}), which is the formula that
we wanted to show.

   From the identity
$$
d(dA)=d(-\theta^{1/2}i_{U}\psi_3)=d(-i_{U}\Psi_3)={\cal L}_{U}\Psi_3=\theta^{1/2}{\cal L}_{U}\psi_3=0,
$$
it follows that $U$ also preserve $\psi_3$. Therefore $U$ is an
isometry of $\overline{J}$ and $\psi_3$. From the results of
\cite{Hitchin2} it is known that a vector field preserving
$\overline{J}$ and $\psi_3$ also preserve $J$. Therefore $U$ is not
only holomorphic, but also hamiltonian. But an holomorphic and
hamiltonian vector field is always \emph{Killing}, that is, it
preserve the metric $g_6$. The $SU(3)$ structure ($J$,
$\overline{J}$, $\psi_3$) is independent of the coordinate $\chi$,
then $U$ can be selected independent of $\chi$ and therefore it
commutes with $V=\partial_{\chi}$. This means that the isometry
group is at least $T^2=U(1)\times U(1)$. It also seen from
(\ref{hamo}) that the variable $\theta$ is the momentum map
corresponding to $U$.

     In conclusion, the monopole system (\ref{monop})
is the (3,1) part of the (\ref{cosn}), and (\ref{monop}) is
satisfied if and only if (\ref{ifo}) is satisfied. \emph{This fact
provides the link between the formalism of the references}
\cite{Apostol} \emph{and} \cite{Kaste}. Therefore we have the
following statement:
\\

\textsf{Corollary} If a ten-dimensional string frame metric in the
presence of the D6-branes is obtained by reduction of an eleven
dimensional background with a $G_2$ holonomy internal manifold which
possess a Killing vector that preserve the $G_2$ structure and is a
warped product over a Kahler 6-dimensional metric, then the original
$G_2$ metric is necessarily toric, i.e, it has at least
$T^2=U(1)\times U(1)$ isometry group.
\\

\textit{The torsion classes of the $SU(3)$ structures}
\\

    From the previous discussion is clear that if a $G_2$ holonomy metric
admits a Kahler reduction along an $U(1)$ isometry $V$ implies the
presence of a new holomorphic isometry $U$ (which is therefore
hamiltonian) such that $[U,V]=0$, together with conditions
(\ref{ifo}) and $F^{(1,1)}=0$. The $G_2$ structure is in this case
\be\lb{lolo} g_{7} = \theta g_{6} +\frac{(d\chi+A)^2}{\theta^2}, \ee
\be\lb{dicol} \Phi= \theta^{3/2} \psi_3 + \overline{J} \wedge e ^z,
\ee \be\lb{dicol2} \ast \Phi= \theta^{1/2}\psi'_3 \wedge e^z+
\frac{1}{2}\theta^{2} \overline{J} \wedge \overline{J}. \ee The next
task is to work out the consequences of the strong supersymmetry
condition $F^{(1,1)}=0$. For this purpose, it is convenient to find
the corresponding torsion classes for our $SU(3)$ structure. In
general, the five torsion classes of a given $SU(3)$ structure are
defined by \cite{Chiozzi} \be\lb{toro1}
d\overline{J}=\frac{3i}{4}(W_1\psi^{*}-\overline{W}_1\psi) +
W_4\wedge \overline{J}+W_3, \ee \be\lb{toro2}
d\psi=\overline{W}_1\overline{J}\wedge\overline{J} + W_2\wedge
\overline{J}+W^{*}_5\wedge \psi, \ee together with the conditions
$$
\overline{J}\wedge W_3=\overline{J}\wedge \overline{J}\wedge
W_2=\psi\wedge W_3=0.
$$
It is seen from the closure of $\overline{J}$ that $W_1=W_4=W_3=0$.
The integrability condition implies that $W_2=0$, which is our
assumption. Therefore the only non vanishing class is $W_5$ and it
is defined by \be\lb{toro3} d\psi=W^{*}_5\wedge \psi. \ee It follows
that $W_5$ is different from zero although $W_4=0$. This is, in
principle, a situation different than those considered in
\cite{Zoupanos}-\cite{Lustzoup}, which arise in heterotic
supersymmetric compactifications with fluxes and condensates. In
order to define $W_5$ it is useful to consider the scaled three form
\be\lb{resca} \Psi=\Psi_3+i\Psi'_3=\theta^{1/2}\psi. \ee As is seen
below (\ref{dim}), $\Psi'_3$ is preserved by $V$ and $d\Psi'_3=0$.
The Kahler assumption implies that $d\Psi_3=0$. This also means that
the transformation $\psi\to\theta^{1/2}\psi$ takes the class $W_5$
to zero, and therefore this class is defined by means of a gradient.
After some calculation it is finally obtain that \be\lb{apos}
d\psi_3=-\frac{3}{2}d \log \theta \wedge \psi'_3,\qquad
d\psi'_3=\frac{3}{2}d^c\log\theta\wedge \psi_3 \ee being $d^c= J d$
defined over the six manifold $N$ \cite{Apostol}. Condition
(\ref{apos}) is equivalent to $F^{(1,1)}=0$ and should be supplied
to (\ref{ifo}) and to the requirement that $U$ is an isometry, in
order to have a Kahler reduction.

\subsubsection{The $G_2$ toric metric and the logarithmic dilaton "evolution"}

   The presence of the new isometry $U$ of the $G_2$ space $Y$ allows
to make a further reduction to a four dimensional $M$ possessing a
complex sympletic structure \cite{Apostol}. The isometry $U$ is
holomorphic and hamiltonian. Therefore there exists a coordinate
system for which the metric $g_6$ and the Kahler form $\overline{J}$
takes the form \be\lb{dik} g_{6}=u \;
d\theta^2+\frac{(d\upsilon+A')^2}{u}+g_{4}(\theta), \ee
\be\lb{dikos}
\overline{J}=\widetilde{J}_1(\theta)+d\theta\wedge(d\upsilon+A'),
\ee being the new Killing vector $U=\partial_{\upsilon}$, and $A'$
certain 1-form. Therefore the Kahler manifold $N$ is locally the
product $N=R_{\theta}\times R_{\upsilon} \times M$ being $M$ certain
four dimensional manifold. The metric $g_{4}(\theta)$ is a metric on
$M$ at each level of constant sets of the coordinate $\theta$.

    It is natural to consider the two forms
$\overline{J}_2=i_{U}\Psi_3$ and $\overline{J}_3=i_{U}\Psi_3'$,
being $\Psi_3$ and $\Psi´_3$ defined in (\ref{resca}). By use of
these definitions and that $\Psi_3$ and $\Psi´_3$ are closed, it
follows that $i_{U}\overline{J}_2=i_{U}\overline{J}_3=0$ and that
$d\overline{J}_2=d\overline{J}_3=0$. The compatibility conditions
(\ref{compat}) implies that \be\lb{lacon} \overline{J}_2\wedge
\overline{J}_2=\overline{J}_3\wedge \overline{J}_3,\qquad
\overline{J}_2\wedge \overline{J}_3=0. \ee These forms are the real
and imaginary part of complex two form
$$
\Omega=i_{\Xi}\Psi,
$$
respectively, being $\Xi$ the holomorphic vector field $\Xi=U-iJU$.
It also holds that $i_{\Xi}\Omega=0$ and that $d\Omega=0$. The form
$\Omega=\overline{J}_2+i\overline{J}_3$ is known as a \emph{complex
sympletic form} on $M$. The complex structure $J$ on the Kahler
6-manifold descends to a complex structure $J_1$ on $M$ which is
obtained from the relation \be\lb{azad}
\overline{J}_2(\cdot,\cdot)=\overline{J}_3(J_1\cdot,\cdot), \ee or
by the relation $g_4(J_1\cdot, \cdot)=\widetilde{J}_1$. Moreover the
equation (\ref{ifo}) implies that \be\lb{ifo2} dA=-\overline{J}_2.
\ee The $U$-invariance of the three forms $\psi_3$ and $\psi_3'$
implies that their general form is \be\lb{trol}
\psi_3=\theta^{-1/2}[\;\overline{J}_2\wedge (d\upsilon+A')+u
\overline{J}_3\wedge d\theta\;], \ee \be\lb{trol2}
\psi_3'=\theta^{-1/2}[\;\overline{J}_3\wedge (d\upsilon+A')-u
\overline{J}_3\wedge d\theta\;], \ee and the Kahler condition
(\ref{apos}) is automatically satisfied for (\ref{trol}) and
(\ref{trol2}). Also the first (\ref{compat}) descends to the
relation \be\lb{compo2} 2\theta \widetilde{J}_1(\theta)\wedge
\widetilde{J}_1(\theta)=u\overline{J}_2\wedge\overline{J}_2
=u\overline{J}_3\wedge\overline{J}_3=u\Omega\wedge
\overline{\Omega}, \ee and that the calibration three form $\Phi$
(\ref{dicol}) is expressed as
$$
\Phi=\widetilde{J}_1(\theta)\wedge (d\chi+A)
+d\theta\wedge (d\upsilon+A')\wedge (d\chi+A)
$$
\be\lb{dale} + \theta\;[\;\overline{J}_2\wedge(d\upsilon+A')+u
\widetilde{J}_1(\theta) \wedge d\theta\;]. \ee Then from
$d\Phi=d\ast\Phi=0$ we obtain the additional equations
\cite{Apostol}
$$
\widetilde{J}_1 ''=-d_{M}d^c_{M} u,\qquad dA'=(d^c_M u) \wedge d\theta
+\widetilde{J}_1',
$$
being $d_M$ defined on $M$ and $d^c_M= J d_M$. In conclusion, and by
denoting now $A=A_2$ and $A'=A_1$, it is concluded that the general
form of the $G_2$ metrics is \cite{Apostol} \be\lb{senio}
g_{7}=\frac{(d\chi+A_2)^2}{\theta^2}+\theta\;[\;u \;
d\theta^2+\frac{(d\upsilon+A_1)^2}{u}+g_{4}(\theta)\;], \ee being
the quantities appearing in (\ref{senio}) defined by the evolution
equations \be\lb{ebol} \widetilde{J}_1''=-d_{M}d^c_{M}u, \ee
\be\lb{compota} 2\theta \widetilde{J}_1(\theta)\wedge
\widetilde{J}_1(\theta)=u\Omega\wedge \overline{\Omega}, \ee and the
forms $A_1$ and $A_2$ are defined on $M\times R_{\theta}$ and $M$
respectively by the equations \be\lb{chon} dA_1=(d^c_M u)\wedge
d\theta+ \widetilde{J}_1',\qquad dA_2=-\overline{J}_2. \ee The
symbol ' denote partial derivation with respect to the parameter
$\theta$.

    We will refer to the metrics (\ref{senio}) as the Apostolov-Salamon
metrics \cite{Apostol}. It is seen from (\ref{dili}) that the
dilaton has a logarithmic behaviour with respect to the "time"
$\theta$\footnote{It is more natural to denote the parameter
$\theta$ as $t$. We did not use this notation in order that the
reader do not get confused with the time coordinate appearing in the
Minkowski metric $g_4$.}. The Kaluza-Klein anzatz (\ref{mert}) can
be rewritten as \be\lb{dilot} g_{11}= \theta g_{10} +
\theta^{-2}(d\chi+A_2)^2 = g_{(3,1)}+ g_7. \ee Our manifolds reduces
in the Type IIA language to a collection of wrapped D6-branes and
the reduced 10-dimensional metric tensor is \be\lb{10} g_{10}=g_6+
\theta^{-1}g_{(3,1)}=u\;
d\theta^2+\frac{(d\upsilon+A_1)^2}{u}+g_{4}(\theta)
+\theta^{-1}g_{(3,1)},\qquad \phi=-\frac{1}{2\alpha}\log(\theta),
\ee and being $A_2$ the potential for the 2 RR form $F=dA_2$. In all
the examples that we can construct by use of the results of this
subsection, the calibration form $\Phi$ given by (\ref{dale}) is not
$L^2$-normalizable, that is, the norm \be\lb{nroo}
||\Phi||=\int_{Y}\Phi\wedge \ast \Phi, \ee badly diverges. Therefore
the scalar mode could be only a real parameter after
compactification to four dimensions.

\section{Explicit toric $G_2$ holonomy metrics}

   In the present section certain solutions of the evolution equations (\ref{ebol}) and
(\ref{compota}) are presented, together with their respective
Apostolov-Salamon metrics. Some of them are known but others are
new. Examples of $G_2$ holonomy manifolds for which for large values
of the evolution parameter $\theta$ tends to a Ricci-flat metric
with $SU(3)$ holonomy are presented. These are asymptotically
Calabi-Yau metrics. Other examples for which this property is not
evident are also constructed. An interesting feature is that a large
class of $G_2$ metrics can be constructed, based on an hyperkahler
4-manifold. There exist a well known family of $G_2$ metrics, the
Bryant-Salamon metrics, that are based on a quaternion Kahler
metric. The metrics of this section are constructed with an
hyperkahler base and they have holonomy exactly $G_2$, even if the
hyperkahler base is flat.

\subsection{Asymptotically Calabi-Yau $G_2$ metrics}

    In searching particular solutions of the evolution equations (\ref{ebol})
and (\ref{compota}) it is important to remark that
$\widetilde{J}_1(\theta)$, $\overline{J}_2$ and $\overline{J}_3$ do
not constitute an hyperkahler structure in general. But a particular
set of solutions of the evolution equations can be found by assuming
that the four manifold $M$ admits an hyperkahler metric $g_{h}$
independent on $\theta$. The closed hyperkahler triplet is also
independent of $\theta$, it will be denoted as $\overline{J}_i$ in
order to do not confuse with $\widetilde{J}_i(\theta)$. Then
equation (\ref{ebol}) is trivially satisfied by an anzatz of the
form \be\lb{monanz}
\widetilde{J}_1(\theta)=\overline{J}_1-\frac{1}{2}d_{M}d^c_{M}
G,\qquad G''=u, \ee being $G$ a function of $\theta$ and of the
coordinates of $M$. It is convenient to introduce the operator
$\textit{M}(G)$ defined through the relation \be\lb{relmon}
(\overline{J}_1-\frac{1}{2}d_{M}d^c_{M} G)^2=
\textit{M}(G)\overline{J}_1\wedge \overline{J}_1. \ee This operator
exists because $\overline{J}_1\wedge \overline{J}_1$ is equal to the
volume form on $M$, and the square of any two form $\omega$ on $M$
is proportional to the volume form, that is
$$
\omega\wedge \omega=A(\omega)\overline{J}_1\wedge \overline{J}_1,
$$
being $A(\omega)$ a function over $M$. In particular by selecting
$\omega=\overline{J}_1-\frac{1}{2}d_{M}d^c_{M} G$ the last relation
gives (\ref{relmon}). The equation (\ref{compota}) can be expressed
as \be\lb{monan} 2\theta \textit{M}(G)=G'', \ee where, as before,
the symbol ' denotes partial derivation with respect to the
evolution parameter $\theta$. The one form $A_1$ is given in this
case by \be\lb{lof} A_1=-\frac{1}{2}d_{M} G'. \ee Equations of the
form (\ref{monan}) have been investigated in the literature
\cite{Bedford}. We see that the left side of (\ref{relmon}) is
explicitly \be\lb{i} \overline{J}_1\wedge \overline{J}_1
-\frac{1}{2}(d_{M}d^c_{M} G) \wedge \overline{J}_1
-\overline{J}_1\wedge \frac{1}{2}(d_{M}d^c_{M} G)
+\frac{1}{4}(d_{M}d^c_{M} G) \wedge (d_{M}d^c_{M} G). \ee From the
last expression together with (\ref{relmon}) it follows that the
operator $\textit{M}(G)$ is not linear in general. An obvious
simplification is obtained when the last term in (\ref{i}) can be
deleted, i.e, when \be\lb{folio} (d_{M}d^c_{M} G)\wedge
(d_{M}d^c_{M} G)=0. \ee Nevertheless, once a solution of
(\ref{monan}) is found by deleting the last term, it should be
checked that such solution is consistent with (\ref{folio}).

   As a ground to the earth, we will consider first the simplest hyperkahler
4-manifold, namely, $R^4$ with its flat metric
$g_4=dx^2+dy^2+dz^2+d\varsigma^2$ and with the hyperkahler triplet
\be\lb{plano} \overline{\widehat{J}}_1=d\varsigma\wedge dy- dz\wedge
dx,\qquad \overline{\widehat{J}}_2=d\varsigma\wedge dx- dy\wedge
dz,\qquad \overline{\widehat{J}}_3=d\varsigma\wedge dz- dx\wedge dy,
\ee which is automatically closed. Let us consider the
simplification (\ref{folio}), then $\textit{M}(G)$ reduce to the
laplacian operator in flat space. If a functional dependence of the
form $G=G(\theta, x,y)$ is selected, then (\ref{monan}) reduces
simply to \be\lb{air} G''+\theta(\partial_{xx}G +
\partial_{yy}G)=2\theta. \ee The separable solutions in
the variable $\theta$ are of the form
$$
G=\frac{1}{3}\theta^3+V(x,y)K(\theta).
$$
By introducing $G=G(\theta, x,y)$ into (\ref{air}) it follows that
$K(\theta)$ and $V(x,y)$ are solutions of the equations \be\lb{m}
K''(\theta)=p\;\theta\; K(\theta),\qquad \partial_{xx}V +
\partial_{yy}V + p\;V=0, \ee being $p$ a parameter. By defining the
$\widetilde{\theta}=\theta/p^{1/3}$ the first of the equations
(\ref{m}) reduce to the Airy equation. The second is reduced to find
eigenfunctions of the two dimensional Laplace operator, which is a
well known problem in electrostatics. For $p>0$ periodical solutions
are obtained and for $p<0$ there will appear exponential solutions.
This solution is consistent with (\ref{folio}).

    A simple example is given by the eigenfunction $V=q\;\sin(p\;x)$, being
$q$ a constant. A solution of the
Airy equation is given by
$$
K=Ai(\widetilde{\theta})=\frac{1}{3}\widetilde{\theta}^{1/2}(J_{1/3}(\tau)
+ J_{-1/3}(\tau)),\qquad \tau=i \frac{2\;\theta^{3/2}}{3\;p^{1/2}}.
$$
Then the function $G$ is
$$
G=\frac{1}{3}\theta^3+ q\;\sin(p\;x) Ai(\frac{\theta}{p^{1/3}}),
$$
From (\ref{lof}) and both equation (\ref{monanz}) it is obtained
\be\lb{kdjd}
A_1=-p\;q Ai(\widetilde{\theta})' \cos(p\;x) dy,\qquad u=\theta (1+ p\;q\; Ai(\widetilde{\theta})\;\sin(p\;x)).
\ee
$$
g_{4}(\theta)=\frac{u}{\theta} (dx^2+ dy^2)+ dz^2+ d\varsigma^2
$$
By defining the new function $H(\theta,x,y)=(1+ p\;q\;
Ai(\widetilde{\theta})\;\sin(p\;x))$  it is obtained the following
$G_2$ holonomy metric \cite{Apostol} \be\lb{airy}
g_{7}=\frac{(d\chi-x dz+y d\varsigma)^2}{\theta^2}
+\frac{(d\upsilon-p\;q Ai(\widetilde{\theta})' \cos(p\;x) dy)^2}{H}
+\theta\;(\;H dx^2+H dy^2+ dz^2+ d\varsigma^2\;) +\theta^2 H \;
d\theta^2. \ee The metric (\ref{airy}) has two parameters $p$ and
$q$ with $p>0$ and three commuting Killing vector fields
$\partial_{\chi}$, $\partial_{\upsilon}$ and $\partial_{\varsigma}$.
The Airy function goes to zero for $\theta\to \infty$ values and
therefore $H$ goes to 1 for large $\theta$. The asymptotic form of
the metric is \be\lb{airy2} g_{7}=\frac{(d\chi-x dz+y
d\varsigma)^2}{\theta^2} +\theta\;(\; dx^2+ dy^2+ dz^2+
d\varsigma^2\;)+\theta^2 \; d\theta^2+d\upsilon^2, \ee and it is
seen that the dependence on $p$ and $q$ have disappeared. It is
immediately seen that (\ref{airy}) is asymptotically of the form
$$
g_7=d\upsilon^2+g_6,
$$
being $g_6$ independent of the coordinate $\upsilon$. Therefore the
holonomy has been reduced from $G_2$ to $SU(3)$, that is, the metric
(\ref{airy}) has asymptotically $SU(3)$ holonomy. It means that the
six dimensional part
$$
g_6=\frac{(d\chi-x dz+y d\varsigma)^2}{\theta^2} +\theta\;(\; dx^2+
dy^2+ dz^2+ d\varsigma^2\;)+\theta^2 \; d\theta^2,
$$
is Calabi-Yau. We do not know if there exist for (\ref{airy}) a
coordinate system for which (\ref{airy}) is asymptotically conical.
Therefore we ignore if (\ref{airy}) is reliable in order to obtain
chiral matter after compactification. We only can say that
(\ref{airy}) is asymptotically Calabi-Yau. In the limit $\theta\to
\infty$ we have the scale invariance
$$
x\longrightarrow \lambda^{3/2} x,\qquad y\longrightarrow \lambda^{3/2} y,\qquad z\longrightarrow \lambda^{3/2} z,\qquad
\varsigma\longrightarrow \lambda^{3/2} \varsigma,
$$
\be\lb{escolo}
\chi\longrightarrow \lambda^3 \chi,\qquad \upsilon\longrightarrow \lambda^4 \upsilon,\qquad
\theta\longrightarrow \lambda \theta,
\ee
which is generated by the homothetic
Killing vector
\be\lb{homos}
D=\frac{3}{2}x\partial_{x}+\frac{3}{2}y\partial_{y}+\frac{3}{2}z\partial_{z}+\frac{3}{2}\varsigma\partial_{\varsigma}
+\theta\partial_{\theta}+
3\chi\partial_{\chi}+4\upsilon\partial_{\upsilon}.
\ee
Instead the full $G_2$ holonomy metric (\ref{airy}) is not invariant under (\ref{homos}), even by a redefinition
of the values of $p$ and $q$.

   As it has been seen in a previous section, after reduction along the coordinate
$\chi$, the dilation has a logarithmic behaviour with respect to the
coordinate $\theta$. An interesting question is which dependence is
obtained by making the IIA reduction along the coordinate
$\upsilon$. The Kaluza-Klein anzatz (\ref{mert}) can be rewritten as
\be\lb{diloton} g_{11}= H^{1/2} g_{10} + H^{-1}(d\upsilon-p\;q
Ai(\widetilde{\theta})' \cos(p\;x) dy)^2. \ee The reduced IIA
background given by
$$
g_{10}=\frac{(d\chi-x dz+y d\varsigma)^2}{H^{1/2}\theta^2}
+\theta\;(\;H^{1/2}dx^2+H^{1/2} dy^2+ H^{-1/2}dz^2+H^{-1/2}
d\varsigma^2\;)
$$
$$
+\theta^2 H^{1/2} \; d\theta^2+ H^{-1/2}g_{(3,1)},
$$
$$
\phi=-\frac{1}{4\alpha}\log(H),
$$
and being $A=-p\;q Ai(\widetilde{\theta})' \cos(p\;x) dy$ the potential for the 2 RR form $F=dA$.
The explicit form of $F$ is
$$
F=p\;q Ai(\widetilde{\theta})' \sin(p\;x) dy\wedge dx-p\;q \widetilde{\theta}
Ai(\widetilde{\theta}) \cos(p\;x) dy\wedge d\widetilde{\theta}
$$
We see that for large $\theta$ the function $H$ goes to one and the
dilaton decreases to zero. The new six dimensional metric will not
satisfy the condition $F^{(1,1)}=0$ but $F^{ab}J_{ab}=0$. Therefore
this new metric is not Kahler.

\subsection{Two parameter $G_2$ metrics with $U(1)\times U(1)\times U(1)$ isometry}

   There are more $G_2$ holonomy metrics that can be constructed by starting with
an hyperkahler 4-manifold and by using the formalism of section
2.1.3. Solutions of the system (\ref{compota}), (\ref{chon}) and
(\ref{ebol}) will be found by assuming the additional condition
$d^c_M u=0$. Then the first equation (\ref{chon}) is solved by
$$
\widetilde{J}_1=(r+s\theta)\;\overline{J}_1,
$$
the evolution equation (\ref{ebol}) is trivially satisfied and
(\ref{compota}) is an algebraic equation for $u$ with solution
$$
u=\theta\;(r+s\theta)^2.
$$
The resulting $G_2$ metric is explicitly \be\lb{metricol} g_7=
\frac{(d\upsilon+A^1)^2}{(r+s\theta)^2}+
\frac{(d\chi+A^2)^2}{\theta^2}
+\theta^2\;(r+s\theta)^2\;d\theta^2+\theta\;(r+s\theta)\;g_{h}, \ee
being $A^1=A^1_i dx^i$ and $A^2=A^2_i dx^i$ one forms defined on the
four manifold $M$ by the equations \be\lb{breath}
dA^1=s\;\overline{J}_1,\qquad dA^2=-\overline{J}_2. \ee The
integrability condition for (\ref{breath}) is satisfied because the
triplet $\overline{J}_i$ of an hyperkahler manifold is always
closed.

   There exist some cases in which a given hyperkahler
metric admit more than one $G_2$ extension. There exist four
dimensional metrics admitting an almost Kahler structure ($I$,
$\overline{J}_0$) compatible with the opposite orientation induced
by the triplet $\overline{J}_i$. A simple example is given by the
flat $R^4$ metric, and hyperkahler examples admitting non integrable
almost Kahler structure are known in the literature \cite{Nurowski}.
For all these cases a new solution of the evolution equations is
obtained with corresponding $G_2$ metric
$$
\widetilde{J}_1=(p+q\theta)\;\overline{J}_0+(r+s\theta)\;\overline{J}_1,\qquad
u=\theta\;[(r+s\theta)^2-(p+q \theta)^2].
$$
\be\lb{metricol3}
g_7= \frac{(d\upsilon+A^1)^2}{(r+s\theta)^2-(p+q\theta)^2}+
\frac{(d\chi+A^2)^2}{\theta^2}
+\theta^2\;[\;(r+s\theta)^2-(p+q\theta)^2\;]d\theta^2
\ee
$$
+\theta\;[\;(r+p)+(q+s)\theta\;]\;g_{h}.
$$
The one forms $A_i$ are defined in this case as
$$
dA^1=s\;\overline{J}_1+p\;\overline{J}_0, \qquad dA^2=-\overline{J}_2.
$$
These equations are well defined due to the closure of
$\overline{J}_i$. Only two of the four parameters $(p, q, r, s)$ are
effective, because two of them can be fixed by a rescale of $\Phi$
and the Killing vector $V$.

    If an hyperkahler metric with no isometries is lifted to a $G_2$
holonomy one by (\ref{metricol}), the isometry group will be exactly
$T^2$. An example of an hyperkahler metric without isometries is the
well known Atiyah-Hitchin metric \cite{Atihich}. But if a larger
isometry group including $T^2$ as a subgroup is desired, then the
hyperkahler basis should have at least one Killing vector. If the
new isometry is not tri-holomorphic, i.e., it do not preserve the
sympletic forms $\overline{J}_i$, then it should preserve at least
$\overline{J}_1$ and $\overline{J}_2$ in order to be an isometry of
the full $G_2$ metric (\ref{metricol}). Because $dA_1\sim
\overline{J}_1$ and $dA_2\sim \overline{J}_2$ it is seen that the
action of this group could merely add a total differential term,
i.e, $A_1\to A_1+ df_1$ and $A_2\to A_2+ df_2$. These terms can be
compensated by a redefinition of the coordinates $\chi$ and
$\upsilon$ and they do not change the local form of the metric.
Nevertheless, as it will be shown below, an isometry preserving
$\overline{J}_1$ and $\overline{J}_2$ is necessarily
tri-holomorphic. If an hyperkahler metric possesses an isometry that
is not tri-holomorphic, then there always exist a coordinate system
$(x, y, z$,$\varsigma)$ for which the distance element take the form
\cite{Boyer} \be\lb{gegodas}
g_{h}=u_z\;[\;e^{u}(dx^2+dy^2)+dz^2\;]+u_{z}^{-1}\;[\;d\varsigma+(u_{x}dy-u_{y}dx)\;]^2,
\ee being $u$ a function of $(x,y,z)$ satisfying the $SU(\infty)$
Toda equation \be\lb{Toda} (e^u)_{zz} + u_{yy} + u_{xx}=0. \ee The
vector field $\partial_{\varsigma}$ is a Killing vector of
(\ref{gegodas}). The metric (\ref{gegodas}) is hyperkahler with
respect to the $\varsigma$-\emph{dependent} hyperkahler triplet
$$
\overline{J}_1=e^u u_z dx\wedge dy+dz\wedge [d\varsigma+(u_{x}dy-u_{y}dx)],
$$
\be\lb{beatit}
\left(\begin{array}{c}
  \overline{J}_2 \\
  \overline{J}_3
\end{array}\right)=e^{u/2}\left(\begin{array}{cc}
  \cos(\frac{\varsigma}{2}) & \sin(\frac{\varsigma}{2}) \\
  \sin(\frac{\varsigma}{2}) & -\cos(\frac{\varsigma}{2})
\end{array}\right)\left(\begin{array}{c}
  \widetilde{\overline{J}}_2 \\
  \widetilde{\overline{J}}_3
\end{array}\right).
\ee In the last expression there have been defined the two forms
$$
\widetilde{\overline{J}}_2=- u_z dz\wedge dy+[d\varsigma+u_{y}dx]\wedge dy,\qquad
\widetilde{\overline{J}}_3= u_z dz\wedge dx+[d\varsigma+u_{x}dy]\wedge dx,
$$
which should not be confused with the hyperkahler forms
$\overline{J}_i$ in (\ref{beatit}). From (\ref{beatit}) it is clear
 that $\partial_{\varsigma}$ preserve $\overline{J}_1$, but the
other two Kahler forms are $\varsigma$ dependent. It means that in
the non-triholomorphic case it is impossible to preserve two of the
three $\overline{J}_i$ without preserving the third. The local form
(\ref{gegodas}) is general, and therefore in four dimensions an
isometry that preserve two of the closed Kahler forms of an
hyperkahler metric is necessarily tri-holomorphic. For any
hyperkahler metric possessing a tri-holomorphic isometry there exist
a coordinate system for which it takes generically the
Gibbons-Hawking form \cite{Gibbhawk} \be\lb{ashgib}
g=V^{-1}(d\varsigma+A^3)^2 + V dx \cdot dx, \ee with a one form
$A^3$ and a function $V$ satisfying the following linear system of
equations \be\lb{Gibb-Hawk} \nabla V=\nabla \times
A^3\;\;\;\;\Longleftrightarrow\;\;\;\;\;\; dV=\ast dA^3, \ee where
the operation $\ast$ is refereed to the three dimensional flat part.
These metrics are hyperkahler with respect to the hyperkahler
triplet
$$
\overline{J}_1=(d\varsigma+A^3)\wedge dx+V dy\wedge dz
$$
\be\lb{transurop}
\overline{J}_2=(d\varsigma+A^3)\wedge dy+V dz\wedge dx
\ee
$$
\overline{J}_3=(d\varsigma+A^3)\wedge dz+V dx\wedge dy
$$
which is clearly $\varsigma$ independent. Only for the
tri-holomorphic case one can expect to obtain $\varsigma$
independent 1-forms $A^1$ and $A^2$, up to a total differential term
that can be managed by a redefinition of the coordinates $\chi$ and
$\upsilon$. The isometry group will be enlarged to $T^3$. In general
an arbitrary isometry group $G$ on $M$ will be enlarged to the total
$G_2$ space if is tri-holomorphic. The total isometry group will be
$T^2\times G$. We will construct examples with larger isometry group
in the next subsection.

\subsection{The flat hyperkahler case}

As before we consider $R^4$ with its flat metric
$g_4=dx^2+dy^2+dz^2+d\varsigma^2$ and with the closed hyperkahler
triplet
$$
\overline{\widehat{J}}_1=d\varsigma\wedge dy- dz\wedge dx,\qquad
\overline{\widehat{J}}_2=d\varsigma\wedge dx- dy\wedge dz,\qquad
\overline{\widehat{J}}_3=d\varsigma\wedge dz- dx\wedge dy.
$$
This innocent looking case is indeed rather rich and instructive. In
principle for any hyperkahler metric there exist three different
$G_2$ holonomy metrics that can be constructed by use of the results
of section 2.1.3. This is due to the fact that in order to integrate
equations (\ref{breath}) we should select a pair of elements of the
triplet $\overline{\widehat{J}}_i$ as $\overline{J}_1$ and
$\overline{J}_2$, and there are three possible choices. But in the
flat case this selection corresponds to a permutation of coordinates
and this ambiguity can be ignored. Equations (\ref{breath}) are
easily integrated for (\ref{plano}) up to a total differential term
and, from (\ref{metricol}), it is directly obtained a $G_2$ metric.
The result is the following
$$
A_1=-s (x dz +  y d\varsigma),
$$
$$
A_2=-y dz-x d\varsigma.
$$
\be\lb{plani}
g_7= \frac{(\;d\upsilon - s (x dz +  y d\varsigma)\;)^2}{( r + s\theta)^2}+
\frac{(d\chi-y dz-x d\varsigma)^2}{\theta^2}
+ \theta^2\;( r + s\theta)^2\;d\theta^2
\ee
$$
+ \theta\;( r + s \theta )(\;dx^2+dy^2+dz^2+d\varsigma^2\;).
$$
If the values $r=0, s=1$ for the parameters are selected the metric
tensor (\ref{plani}) reduce to \be\lb{plani2} g_7=
\frac{(d\upsilon-x dz + y d\varsigma)^2}{\theta^2}+ \frac{(d\chi-y
dz-x d\varsigma)^2}{\theta^2} +\theta^4\;d\theta^2
+\theta^2\;(\;dx^2+dy^2+dz^2+d\varsigma^2\;). \ee The metrics
(\ref{plani2}) have been already obtained in the physical literature
\cite{Stele}. They have been constructed by use of oxidation methods
in 11 dimensional supergravity, by starting with a domain wall
solution in five dimensions of the form
$$
g_5=H^{4/3}(\;dx^2+dy^2+dz^2+d\varsigma^2\;)+H^{16/3}d\theta^2,
$$
being $a=1,..,4$. By use of oxidation rules one obtains a 11
dimensional background $g_{11}=g_{(3,1)}+g_7$ with a 7-dimensional
metric of the form \be\lb{oxo} g_7= \frac{(d\upsilon-x dz + y
d\varsigma)^2}{H^2}+ \frac{(d\chi-y dz-x d\varsigma)^2}{H^2}
+H^4\;d\theta^2 +H^2\;(\;dx^2+dy^2+dz^2+d\varsigma^2\;). \ee The
last metric reduce to (\ref{plani2}) by selecting $H=\theta$. It can
be shown that (\ref{oxo}) arise from an $SO(5)$ invariant $G_2$
metric by contraction on the isometry group \cite{Stele}. The
isometry corresponding to (\ref{oxo}) and (\ref{plani}) is the same
$SU(2)$ group that acts linearly on the coordinates
$(x,y,z,\varsigma)$ on $M$ and which preserve simultaneously the two
forms $\overline{J}_1$ and $\overline{J}_2$. The nilpotent
6-dimensional algebra is the complexification of the 3-dimensional
ur-Heisenberg algebra. The $SU(2)$ action on $M$ adds to $A_1$ and
$A_2$ a total differential term that can be absorbed by a
redefinition of the coordinates $\chi$ and $\upsilon$.  For
instance, by general translation of the form \be\lb{sip}
x\longrightarrow x+\alpha_1,\qquad y\longrightarrow
y+\alpha_2,\qquad z\longrightarrow z+\alpha_3,\qquad
\varsigma\longrightarrow \varsigma+\alpha_4, \ee preserves
$\overline{J}_1$ and $\overline{J}_2$ but not the one forms $A_1$
and $A_2$. Nevertheless the effect of (\ref{sip}) can be compensated
by a transformation of the form \be\lb{sop} \upsilon\longrightarrow
\upsilon +\alpha_5 + \alpha_1 z -\alpha_2 \varsigma ,\qquad
\chi\longrightarrow \chi +\alpha_6 + \alpha_2 z +\alpha_1 \varsigma.
\ee More general $SU(2)$ transformations on $M$ preserving
$\overline{J}_1$ and $\overline{J}_2$  will also be absorbed by a
redefinition of the coordinates $\chi$ and $\upsilon$. For the
metric (\ref{plani2}) we also have the scale invariance
$$
x\longrightarrow \lambda x,\qquad y\longrightarrow \lambda y,\qquad z\longrightarrow \lambda z,\qquad
\varsigma\longrightarrow \lambda \varsigma,
$$
$$
\chi\longrightarrow \lambda^4 \chi,\qquad \upsilon\longrightarrow \lambda^4 \upsilon,\qquad
\theta\longrightarrow \lambda^2 \theta,
$$
which is generated by the homothetic
Killing vector
\be\lb{homo}
D=2x\partial_{x}+2y\partial_{y}+2z\partial_{z}+2\varsigma\partial_{\varsigma}+\theta\partial_{\theta}+
4\chi\partial_{\chi}+4\upsilon\partial_{\upsilon}.
\ee
It can be shown by explicit calculation of the curvature
tensor that (\ref{oxo}) is irreducible and has holonomy
exactly $G_2$, even in the subcase given in (\ref{plani2}).

   But more metrics can be found starting with the flat metric in $R^4$.
This metric is also hyperkahler with respect to the opposite
orientation triplet \be\lb{opo} \overline{J}_1'=d\varsigma\wedge dy+
dz\wedge dx,\qquad \overline{J}_2'=d\varsigma\wedge dx+ dy\wedge
dz,\qquad \overline{J}_3'=d\varsigma\wedge dz+ dx\wedge dy, \ee and
the three one forms \be\lb{joe} A_1'=x dz+y d\varsigma,\qquad A_2'=z
dy+ x d\varsigma, \qquad A_3'=y dx + z d\varsigma, \ee satisfy
$dA_i'=\overline{J}_i'$. By denoting
$\overline{J}_0=\overline{J}_1'$, then $g_4$ and $\overline{J}_0$
are a Kahler structure compatible with the opposite orientation
defined by $\overline{J}_i$. Therefore from (\ref{metricol3}) we
obtain the new $G_2$ metric
$$
g_7= \frac{[\;d\upsilon-(s-q) x dz + (s+q)y d\varsigma\;]^2}{(r+s\theta)^2-(p+q\theta)^2}+
\frac{(\;d\chi-y dz-x d\varsigma\;)^2}{\theta^2}
$$
\be\lb{metricol4}
+\theta\;(\;r+p+(q+s)\theta\;)\;[\;\theta\;(\;r-p+(s-q)\theta)\;d\theta^2
+\;dx^2+dy^2+dz^2+d\varsigma^2\;].
\ee
For the value $s=1$ and all the remaining parameters equal to zero, the last metric
reduce to (\ref{plani2}). The metric (\ref{metricol4}) is not defined for
the values $p=r$ and $s=q$. If $s=q$ but $p\neq q$, then the expression
of the (\ref{metricol4}) is simplified as
$$
g_7= \frac{(d\upsilon + 2sy d\varsigma )^2}{f(\theta)}+
\frac{(d\chi-y dz-x d\varsigma)^2}{\theta^2}
+\theta^2 f(\theta) d\theta^2
$$
\be\lb{metricol4a} +\frac{\theta
f(\theta)}{(r-p)}(dx^2+dy^2+dz^2+d\varsigma^2). \ee being
$f(\theta)$ now a linear function of $\theta$ defined by
$f(\theta)=(r-p)(2s \theta +r+p)$. It can be immediately checked
that for $s=-q$ it is obtained again the metric (\ref{metricol4a})
up to a redefinition of the parameters and coordinates. The metric
(\ref{metricol4}) is invariant under the linear action on the
coordinates on the manifold preserving the two forms
$s\overline{J}_1+p\overline{J}_0$ and $\overline{J}_2$. If we
instead choose $\overline{J}_2'$ or $\overline{J}_3'$ as
$\overline{J}_0$, then we will find two more $G_2$ metrics and a
similar analysis of the isometries and parameters could be done. We
will not write the explicit expressions, but the method to find them
is completely analogous to those described above. In all the cases
presented here, there is a $U(1)\times U(1)\times U(1)$ subgroup of
isometries generated by shifts on $\chi$, $\upsilon$ and
$\varsigma$.

\subsection{The stringy cosmic string case}

  After this warm up, more complicated examples can be analyzed. Let us consider
hyperkahler metrics with two commuting Killing vectors, one of which
is tri-holomorphic and other that is not. The presence of a
tri-holomorphic Killing vector implies that the existence of a
coordinate system for which such metrics will take the
Gibbons-Hawking form (\ref{ashgib}). It is convenient to write the
flat $3$-dimensional part in cylindrical coordinates
$(\eta,\rho,\varphi)$, the result is
$$
g_f = dx^2 + dy^2 + dz^2 = d\eta^2 + d\rho^2 + \rho^2 d\varphi^2.
$$
Then the subclass of the metrics (\ref{ashgib}) for which $\varphi$
is a Killing vector take in cylindrical coordinates the following
form \be\lb{queseyo1}
g_h=V(d\eta^2+d\rho^2+\rho^2d\phi^2)+V^{-1}(d\varsigma + A d\phi)^2.
\ee The functions $V$ and $A$ for (\ref{queseyo1}) satisfy the
linear system the linear system \be\lb{ward} \rho
V_{\eta}=A_{\rho},\qquad \rho V_{\rho}=-A_{\eta}, \ee which in
particular implies that $V$ satisfy the Ward integrability condition
\be\lb{ord} (\rho V_{\eta})_{\eta}+(\rho V_{\rho})_{\rho}=0. \ee The
Killing vector $\partial_{\varsigma}$ and $\partial_{\phi}$ are
commuting and therefore there is an $T^2$ action over the
4-manifold. Nevertheless the former isometry is not tri-holomorphic
and therefore will not be an isometry of the $G_2$ extension.

   A simple example of (\ref{queseyo1}) occurs when $V$ is independent
on the coordinate $\eta$. The Ward equation gives
$V_{\eta}=\log(\rho)$ and the distance element is \be\lb{cordoba}
g_h=\log(\rho)(d\eta^2+d\rho^2+\rho^2d\phi^2)+\frac{1}{\log(\rho)}(d\varsigma
- \eta d\phi)^2. \ee The metric (\ref{cordoba}) appears in many
contexts in physics as, for instance, in the study of stringy cosmic
strings \cite{Vafayau}. It is the asymptotic form of the ALG
gravitational instantons found in \cite{ALG}. Also it describe the
single matter hypermultiplet target metric for type IIA superstrings
compactified on a Calabi-Yau threefold when supergravity and
D-instanton effects are absent \cite{Vafa}. Our strategy will be, as
before, to express the hyperkahler triplet (\ref{transurop})
corresponding to the metric (\ref{cordoba}) as the differential of
certain one forms $A_i$, and to construct the corresponding $G_2$
metrics by use of the formulas (\ref{breath}) and (\ref{metricol}).
After certain calculation it is found that the triplet can be
expressed in cylindrical coordinates as
$\overline{\widehat{J}}_i=dA_i$ being $A_i$ defined by
$$
A_1=\rho \cos(\phi)d\varsigma-\rho\eta \cos(\phi)d\phi + \sin(\phi)(\rho \log(\rho)-\rho)d\eta,
$$
\be\lb{into}
A_2=\rho \sin(\phi)d\varsigma + \rho\eta \sin(\phi)d\phi+\cos(\phi)(\rho \log(\rho)-\rho)d\eta,
\ee
$$
A_3=\eta d\varsigma+(\eta^2+\rho \log(\rho)-\rho)d\phi.
$$
With the help of the last expressions and (\ref{metricol}) and
(\ref{breath}) we can construct three, in principle different, $G_2$
holonomy metrics, depending on which pair of 2-forms
$\overline{\widehat{J}}_i$ will play the role of $\overline{J}_1$
and $\overline{J}_2$. For instance, by selecting
$\overline{\widehat{J}}_1$ and $\overline{\widehat{J}}_3$ as
$\overline{J}_1$ and $\overline{J}_2$, and by using (\ref{into}),
(\ref{metricol}) and (\ref{breath}) it is obtained the following
$G_2$ metric
$$
g_7= \frac{[\;d\upsilon +\rho \cos(\phi)d\varsigma-\rho\eta \cos(\phi)d\phi + \sin(\phi)(\rho \log(\rho)-\rho)d\eta\;]^2}{(r+s\theta)^2}
+\theta^2(r+s\theta)^2 d\theta^2
$$
\be\lb{planilo}
+\theta (r+s\theta)[\;\log(\rho)(d\eta^2+d\rho^2+\rho^2d\phi^2)+\frac{1}{\log(\rho)}(d\varsigma - \eta d\phi)^2\;]
\ee
$$
+\frac{[\;d\chi-s \eta d\varsigma-s(\eta^2+\rho \log(\rho)-\rho)d\phi\;]^2}{\theta^2}.
$$
We will not write the expressions of the other two $G_2$ metrics, but the procedure
for constructing them is completely analogous. The Killing vector fields $\partial_{\eta}$ and $\partial_{\phi}$
are not tri-holomorphic and therefore are not Killing vectors of the full $G_2$ metric.

       More examples can be constructed
by finding solutions that depends on the coordinate $\eta$. For
instance, we can consider solutions of the form \be\lb{tobo} V=a
\log(\rho)+ b \eta +c \log(\frac{\eta+\sqrt{\rho^2+\eta^2}}{\rho}),
\ee which give rise to Taub-Nut solutions, and also solutions like
\be\lb{egohan} V=a \log(\rho)+\frac{1}{2}( b+c/\epsilon
)\log(\frac{\eta-\epsilon+\sqrt{\rho^2+(\eta-\epsilon)^2}}{\rho})
+\frac{1}{2}( b-c/\epsilon
)\log(\frac{\eta+\epsilon+\sqrt{\rho^2+(\eta+\epsilon)^2}}{\rho}),
\ee with $\epsilon^2=\pm 1$. The case corresponding to minus sign
corresponds to the potential for the axially symmetric circle of
charge, while the other case corresponds to two sources on the axis
of symmetry. This are known as Eguchi-Hanson I and II, the second is
complete but the first is not. Finally one can consider multi-center
Gibbons-Hawking metrics as well, which in the case of $N$ coincident
centers is related to a Wu-Yang solution. To find the $G_2$ metrics
corresponding to all this cases is straightforward, but the
expressions are rather cumbersome and we will not write them
explicitly.

\subsection{Half-flat associated metrics}

   It is not difficult to see that there exist a coordinate system
for which the metrics (\ref{metricol3}) takes the form \be\lb{to}
g_7=d\tau^2+ g_{6}(\tau), \ee being $g_6(\tau)$ a six dimensional
metric depending on $\tau$ as an evolution parameter. In fact, by
introducing the new variable $\tau$ defined by \be\lb{mu}
\theta^2\;[\;(r+s\theta)^2-(p+q\theta)^2\;]\;d\theta^2=d\tau^2, \ee
it is seen that (\ref{metricol3}) takes the desired form. Therefore
these $G_2$ holonomy metrics are a wrapped product $Y=I_{\tau}\times
N'$ being $I$ a real interval. The coordinate $\tau$ is just a
function of $\theta$ and is given by \be\lb{inot} \tau=\int
\theta\;[\;(r+s\theta)^2-(p+q\theta)^2\;]^{1/2}\;d\theta. \ee The
point is that $g_{6}(\tau)$ is a half-flat metric on any
hypersurface $Y_{\tau}$ for which $\tau$ has constant value. Indeed
the $G_2$ structure can be decomposed as \be\lb{jom}
\Phi=\widehat{J}\wedge d\tau + \widehat{\psi}_3, \ee \be\lb{joma}
\ast \Phi=\widehat{\psi}_3'\wedge d\tau
+\frac{1}{2}\widehat{J}\wedge \widehat{J}, \ee where we have defined
\be\lb{jj} \widehat{J}=z^{1/2}\;\overline{J}_3+z^{-1/2}(d\upsilon +
A_1)\wedge(d\chi + A_2), \ee \be\lb{daj}
\widehat{\psi}_3=z^{-1/2}\;\widetilde{J}_1\wedge (d\chi + A_2)+
\theta\overline{J}_2\wedge (d\upsilon + A_1), \ee \be\lb{daj2}
\widehat{\psi}_3'=\theta\; z^{-1/2}\overline{J}_2\wedge (d\chi +
A_2)-\theta^2 z^{1/2} \widetilde{J}_1\wedge (d\upsilon + A_1), \ee
and $z=\theta^2\;[\;(r+s\theta)^2-(p+q\theta)^2\;]$. Then the $G_2$
holonomy conditions $d\Phi=d\ast\Phi=0$ are
$$
d\Phi=d\widehat{\psi}_3 +(d\widehat{J}-\frac{\partial \widehat{\psi}_3}{\partial \tau})\wedge d\tau=0,
$$
$$
d\ast\Phi=\widehat{J}\wedge d\widehat{J}+
(d \widehat{\psi}_3+\widehat{J}\wedge \frac{\partial \widehat{J}}{\partial \tau})\wedge d\tau=0.
$$
The last equations are satisfied if and only if \be\lb{half-flat}
d\widehat{\psi}_3=\widehat{J}\wedge d\widehat{J}=0 \ee for every
fixed value of $\tau$, and \be\lb{ovo} \frac{\partial
\widehat{\psi}_3}{\partial \tau}=d\widehat{J},\qquad
\widehat{J}\wedge \frac{\partial \widehat{J}}{\partial \tau} =- d
\widehat{\psi}_3. \ee The flow equations (\ref{ovo}) were considered
by Hitchin and are related to certain Hamiltonian system
\cite{Hitchin}, \cite{Apostol}. Equations (\ref{half-flat}) implies
that for every constant value $\tau$ the metric $g_6$ together with
$\widehat{J}$, $\widehat{\psi}_3$ and $\widehat{\psi}_3'$ form a
half-flat or half-integrable structure \cite{Chiozzi}. It means that
the only non vanishing torsion classes defined in (\ref{toro1}) and
(\ref{toro2}) are $W_1$, $W_2$ and $W_3$. A more precise description
is as follows. Let us consider the decomposition $W_1=W_1^{+}+W_1^-$
and $W_2=W_2^{+}+W_2^-$ given by
$$
d\widehat{\psi}_3 \wedge \overline{J}=\widehat{\psi}_3 \wedge d\overline{J}=
W_1^+ \overline{J}\wedge \overline{J} \wedge \overline{J},
$$
$$
d\widehat{\psi}_3' \wedge \overline{J}=\widehat{\psi}_3' \wedge d\overline{J}=
W_1^- \overline{J}\wedge \overline{J} \wedge \overline{J},
$$
$$
(d\widehat{\psi}_3)^{(2,2)}=W_1^+\overline{J}\wedge \overline{J}+W_2^+\wedge \overline{J},
$$
$$
(d\widehat{\psi}_3')^{(2,2)}=W_1^-\overline{J}\wedge \overline{J}+W_2^-\wedge \overline{J}.
$$
Then for a half-flat manifold $W_1^+=W_2^+=W_4=W_5=0$ and therefore
the intrinsic torsion take values in $W_1^-\oplus W_2^-\oplus W_3$.
It is direct to construct the half-flat metrics corresponding to the
$G_2$ holonomy metrics presented of this section. For instance for
(\ref{plani}) and (\ref{planilog}) we obtain \be\lb{planig} g_6=
\frac{(\;d\upsilon - s (x dz +  y d\varsigma)\;)^2}{(r+s a)^2}+
\frac{(d\chi-y dz-x d\varsigma)^2}{a^2} \ee
$$
+ a\;(r+s a)(\;dx^2+dy^2+dz^2+d\varsigma^2\;),
$$
and
$$
g_6= \frac{[\;d\upsilon +\rho \cos(\phi)d\varsigma-\rho\eta \cos(\phi)d\phi
+ \sin(\phi)(\rho \log(\rho)-\rho)d\eta\;]^2}{(r+s a)^2}
$$
\be\lb{planilog}
+a\;b[\;\log(\rho)(d\eta^2+d\rho^2+\rho^2d\phi^2)+\frac{1}{\log(\rho)}(d\varsigma - \eta d\phi)^2\;]
\ee
$$
+\frac{[\;d\chi-s \eta d\varsigma-s(\eta^2+\rho \log(\rho)-\rho)d\phi\;]^2}{a^2}.
$$
respectively. Here $a$ was a function of $\tau$ in the original
$G_2$ metric, but because we are considering constant $\tau$ it plays
the role as a parameter in the six dimensional metric $g_6$.

\section{D6 brane backgrounds and their $\gamma$-deformations}

  It is convenient to describe in more detail the $SL(2,R)$ solution
generating technique sketched in the introduction and developed in
\cite{Lunin}. One usually starts with a solution of the eleven
dimensional supergravity with $U(1)\times U(1)\times U(1)$ isometry.
Such solution can be written in the generic form \be\lb{as}
g_{11}=\Delta^{1/3}M_{ab}D\alpha_a
D\alpha_b+\Delta^{-1/6}\widetilde{g}_{\mu\nu}dx^{\mu}dx^{\nu}, \ee
$$
C_3=C D\alpha_1\wedge D\alpha_2\wedge D\alpha_3+C_{1(ab)} \wedge
D\alpha_a\wedge D\alpha_b+ C_{2(a)}D\alpha_a+C_{(3)},
$$
with the indices a,b=1,2,3 are associated to three isometries
$\alpha_1=\upsilon$, $\alpha_2=\chi$ and $\alpha_3=\varsigma$ and
the Greek indices $\mu$, $\nu$ run over the remaining eight
dimensional coordinates. Here $D\alpha_i=d\alpha_i+A_i$, being $A_i$
a triplet of one forms defined over the remaining eight dimensional
manifold. In the present framework there is a manifest $SL(3,R)$
symmetry which acts on the coordinates $\alpha_i$ as \be\lb{sl3}
\left(\begin{array}{c}
  \alpha_1 \\
  \alpha_2 \\
  \alpha_3
\end{array}\right)'=(\Lambda^{T})^{-1}\left(\begin{array}{c}
  \alpha_1 \\
  \alpha_2 \\
  \alpha_3
\end{array}\right),
\ee
and over the field tensors as $M$ and $A_i$ as
\be\lb{sl2} M'=\Lambda M \Lambda^T
\qquad \left(\begin{array}{c}
  A_1 \\
  A_2 \\
  A_3
\end{array}\right)'=(\Lambda^{T})^{-1}\left(\begin{array}{c}
  A_1 \\
  A_2 \\
  A_3
\end{array}\right).
\ee
$$
\left(\begin{array}{c}
{C}_{23\mu}\\{C}_{31\mu}\\{ C}_{12\mu}
\end{array}\right)\rightarrow
(\Lambda^T)^{-1}
\left(\begin{array}{c}
{C}_{23\mu}\\{C}_{31\mu}\\{  C}_{12\mu}
\end{array}\right),\qquad
\left(\begin{array}{c}
{C}_{1\mu\nu}\\{ C}_{2\mu\nu}\\C_{3\mu\nu}
\end{array}\right)\rightarrow \Lambda
\left(\begin{array}{c}
{  C}_{1\mu\nu}\\{  C}_{2\mu\nu}\\{C}_{3\mu\nu}
\end{array}\right).
$$
The full isometry group of 11-dimensional supergravity compactified
on a three torus is $SL(3,R)\times SL(2,R)$. The $SL(3,R)$ group
leaves the background (\ref{as}) unaltered. Following \cite{Lunin}
we will deform these $T^3$ invariant backgrounds by an element of
$SL(2,R)$. The strategy for deforming (\ref{as}) is to use an
$SL(2,R)$ transformation described as follows \cite{Cremer}. Let us
define the complex parameter $\tau=C+i \Delta^{1/2}$. Under the
$SL(2,R)$ action $\tau$ is transformed as \be\lb{taum}
\tau\longrightarrow \frac{a \tau + b}{c \tau +d}; \qquad
\Lambda=\left(\begin{array}{cc}
  a & b \\
  c & d
\end{array}\right)\in SL(2,R).
\ee The eight dimensional metric $g_{\mu\nu}$ and the tensor $C_2$
are instead invariant. The tensor $C_{(1)ab\mu}$ and $A_{\mu}^a$
form a doublet in similar way that the RR and NSNS two form fields
do in IIB supergravity, their transformation law is given by
\be\lb{doblete} B^a=\left(\begin{array}{c}
  2 A^a \\
  -\epsilon^{abc}C_{(1)bc}
\end{array}\right),\qquad B^a\longrightarrow \Lambda^{-T}B^a
\ee
The field strenght $C_3$ also form a doublet with its magnetic dual with
consequent transformation law
\be\lb{doblete}
H=\left(\begin{array}{c}
  F_4 \\
  \Delta^{-1/2}\ast_{8} F_{4} + C_{(0)}F_{4}
\end{array}\right),
\qquad H\longrightarrow \Lambda^{-T}H,
\ee
being the Hodge operation taken with respect to the eight dimensional
metric $g$. As we discussed
in the introduction, this transformation
deform the original metric (\ref{as}) and
the deformed metric will be
regular only with elements
of the form \cite{Lunin}
\be\lb{gamo}
\Lambda=\left(\begin{array}{cc}
  1 & 0 \\
  \gamma & 1
\end{array}\right)\in SL(2,R),
\ee which constitute a subgroup called $\gamma$-transformations. We
will be concerned with such transformation in the following.

   If the fields $C$, $C_1$ and $C_2$ are zero, it follows that $A^i$
and $\widetilde{g}_{\mu\nu}$ are unchanged by a
$\gamma$-transformation and $C_1$ and $C_2$ remains zero. The
deformation then give the new fields \be\lb{inge}
\Delta'=G^2\Delta,\qquad C'=-\gamma G \Delta,\qquad
G=\frac{1}{1+\gamma^2 \Delta}. \ee By inspection of the
transformation rule (\ref{doblete}) it follows that \be\lb{gener}
F_4'=F_4-\gamma\Delta^{-1/2}\ast_{8} F_{4}-\gamma d(G\Delta
D\alpha_1\wedge D\alpha_2 \wedge D\alpha_3). \ee The
$\gamma$-deformed eleven dimensional metric results \cite{Cremer}
\be\lb{ede} g_{11}=G^{-1/3}(G\Delta^{1/3}M_{ab}D\alpha_a
D\alpha_b+\Delta^{-1/6}\widetilde{g}_{\mu\nu}dx^{\mu}dx^{\nu}). \ee
 Note that if the initial four form $F_4$ was zero, then from the
last term in (\ref{gener}) a non trivial flux is obtained.

  The gamma-deformation procedure can be applied to a generic solution of the eleven
dimensional supergravity over a manifold of the form $M_{3,1}\times
Y$, being $M_{3,1}$ the flat Minkowski four manifold and $Y$ a
manifold with holonomy in $G_2$ and $T^3$ isometry. In particular it
can be applied for the $G_2$ metrics presented along this work. The
general solution is of the form \be\lb{om} g_{11}=g_{(3,1)}+g_{7}.
\ee being $g_{(3,1)}$ the flat Minkowski metric in four dimensions.
The form $C_3$ vanish identically and the only non trivial field is
the graviton.The local form of the metric will be \be\lb{pipino}
g_{7}=\frac{(d\chi+A_2)^2}{\theta^2}+\theta\;[\;u \;
d\theta^2+\frac{(d\upsilon+A_1)^2}{u} +
\frac{(d\varsigma+A_3)^2}{W}\;]+ g_{3}(\theta), \ee being $W$ a new
function, and being all the quantities appearing in (\ref{pipino})
independent on the coordinates $\chi$, $\upsilon$ and $\varsigma$.
The metric $g_{3}(\theta)$ is a three dimensional metric at every
constant $\theta$ level surface. The reader can check that
\emph{all} the $G_2$ holonomy metrics presented along this work can
be expressed as (\ref{pipino}). The components of the 7-metric are
$$
g_{\upsilon\upsilon}=\frac{\theta}{u},\qquad g_{\chi\chi}=\frac{1}{\theta^2},\qquad
g_{\varsigma\varsigma}=\frac{\theta}{W},\qquad g_{\theta\theta}=\theta u,
$$
$$
g_{\upsilon\varsigma}=\frac{\theta \; A^1_{\varsigma}}{u},\qquad
g_{\upsilon x^i}=\frac{\theta\;A^1_{x^i}}{u},
$$
\be\lb{explotsa}
g_{\chi\varsigma}=\frac{A^2_{\varsigma}}{\theta^2},\qquad
g_{\chi x^i}=\frac{A^2_{x^i}}{\theta^2},\qquad
g_{\varsigma x^i}=\frac{\theta \; A^3_{x^i}}{W},\qquad
\ee
$$
g_{x^i x^j}= g_{3 x^i x^j}+\frac{\theta\; A^1_{x^i} \; A^{1}_{x^j}}{u}
+\frac{A^2_{x^i} \; A^{2}_{x^j}}{\theta^2}+\frac{\theta\;A^3_{x^i}\; A^{3}_{x^j}}{W},
$$
$$
g_{x^i x^i}= g_{3 x^i x^i}+ \frac{\theta\;(A^1_{x^i})^2}{u}
+\frac{(A^2_{x^i})^2}{\theta^2}+\frac{\theta\; (A^3_{x^i})^2}{W},
$$
and the remaining components are zero.

    The task to write the background (\ref{om}) in the $SL(3,R)$ manifest
form (\ref{as}) presents no difficulties for the metrics
(\ref{explotsa}). By defining the coordinates $\alpha_1=\upsilon$,
$\alpha_2=\chi$ and $\alpha_3=\varsigma$ the metric (\ref{om}) can
be expressed as
$$
g_{11}=g_{(3,1)}+\Omega_{ab}(d\alpha_a+A^a)(d\alpha_b+A^b)+h,
$$
where the matrix $\Omega_{ab}$ and the eight dimensional metric $h$
are defined by
$$
\Omega_{ab}=\left[\begin{array}{ccc}
  \frac{\theta}{u} & 0 & 0 \\
  0 & \theta^{-2} & 0 \\
  0 & 0 & \frac{\theta}{W}
\end{array}\right],\qquad
\det(\Omega)=\frac{1}{u\; W},
$$
$$
h=g_{3}(\theta)+\theta\;\;u \; d\theta^2+ g_{(3,1)}.
$$
Therefore the unit determinant matrix $M_{ab}$ and the scalar $\Delta$ are given by
$$
M_{ab}=\frac{\Omega_{ab}}{\det(\Omega)},\qquad \Delta=\det(\Omega)^3=\frac{1}{u^3\; W^3}.
$$
Introducing the covariant derivative $D\alpha_i=d\alpha_i+A^i$
gives, after making the identification
$\widetilde{g}=\Delta^{1/6}h$, the desired $SL(3,R)$ form
(\ref{as}). The deformation technique gives the following
deformation \be\lb{inges} \Delta'=G^2\Delta,\qquad C'=-\gamma G
\Delta,\qquad G=\frac{1}{1+\gamma^2 \Delta} \ee
$$
F_4'=-\gamma d(G\Delta D\alpha_1\wedge D\alpha_2
\wedge D\alpha_3).
$$
and the new metric tensor \be\lb{edes}
g_{11}=G^{-1/3}(G\Delta^{1/3}M_{ab}D\alpha_a
D\alpha_b+\Delta^{-1/6}\widetilde{g}_{\mu\nu}dx^{\mu}dx^{\nu}) \ee
The new solution include a flux term $F_4'$ that was absent in the
starting background (\ref{as}).

  Also the presence of an isometry in the 11-dimensional background
allows to find a IIA background by reduction along the Killing
vector. If after this reduction a new isometry is preserved, then
T-duality rules can be used in order to construct IIB backgrounds.
In principle there are six possible reductions that can be done,
depending on which pair of isometries is choose to make a reduction
along a circle and a T-duality afterwards. In order to perform the
IIA reduction the $T^3$ part of the metric should be decomposed as
\be\lb{decom} M_{ab}D\alpha_a D\alpha_b=e^{-2\phi/3}h_{mn}D\alpha_m
D\alpha_n + e^{4\phi/3}(D\alpha_3 + N_m D\alpha_m)^2, \ee with the
indices $m,n=1,2$. The field $\phi$ will be related to the dilaton
of the reduced theory. By reducing along the isometry and, after
that, making a T-duality along one of the remaining isometries, say
$\alpha_1$, a IIB supergravity solution will be obtained. The final
result can be found elsewhere, for instance in $SL(3,R)$ form in
\cite{Lunin} . In the case corresponding to (\ref{om}) and
(\ref{pipino}), the $T^3$ metric can be decomposed as in
(\ref{decom}) easily, and it is seen that the quantities $N_m$ are
zero and the term $e^{4\phi/3}$ associated to the dilaton will be
one of diagonal elements of $\Omega_{ab}$, associated to our choice
of the pair of isometries. The resulting background is
$$
g_{IIB}=\frac{1}{h_{11}} \left[ {1 \over
\sqrt{\Delta} } (D\alpha^1)^2 +\sqrt{\Delta}
(D\alpha^2)^2 \right]
+e^{2\phi/3}\widetilde{g}_{\mu\nu}dx^\mu dx^\nu,
$$
\be\lb{sumo}
B=\frac{h_{12}}{h_{11}}D\alpha^1 \wedge D\alpha^2,\qquad
C^{(2)}=- D\alpha^1\wedge
A^3_\mu dx^\mu,\qquad C^{(4)}=0
\ee
$$
e^{2\Phi}=\frac{e^{2\phi}}{h_{11}},\qquad C^{(0)}=0.
$$
It is worthy to recall that ,in general, a IIB solution contains
more tensor fields than those appearing in (\ref{sumo}). This
tensors have well defined transformation properties under the
$SL(3,R)$ action. But these fields are zero in our case. Instead
under the $SL(2,R)$ group that generates new solutions, we have a
complex parameter $C+i\sqrt{\Delta}$ transforming as a $\tau$
parameter. There is a four form $F_4$ defined in terms of certain
field $C_{\mu\nu\alpha}$ appearing in the general expression
$C_{(4)}$. This form transforms into its magnetic dual in eight
dimensions as in (\ref{doblete}). Although this field is zero in our
case, the $SL(R,2)$ transformation induce a non trivial $F_4$ term.
We have checked that this IIB deformed background is indeed the
background obtained by reduction of the deformed 11-supergravity
solution given in (\ref{edes}), the final result is
$$
g_{IIB}=\frac{1}{h_{11}} \left[ {1 \over
\sqrt{\Delta'} } (D\alpha^1-C D\alpha_2)^2 +\sqrt{\Delta'}
(D\alpha^2)^2 \right]
+e^{2\phi/3}\widetilde{g}_{\mu\nu}dx^\mu dx^\nu,
$$
\be\lb{sumo}
B=\frac{h_{12}}{h_{11}}D\alpha^1 \wedge D\alpha^2,\qquad
C^{(2)}=- D\alpha^1\wedge
A^3_\mu dx^\mu,\qquad C^{(4)}=0
\ee
$$
e^{2\Phi}=\frac{e^{2\phi}}{h_{11}},\qquad C^{(0)}=0.
$$
being $C'$ and $\Delta'$ defined in (\ref{inges}).

 It is important to recall that a D-brane on the original background
that is invariant under both $U(1)$ symmetries, will be left invariant
under the action of (\ref{tau}). Therefore the new generated background will
contain also a D-brane. It has been conjectured  \cite{Lunin} that
if the original brane gave rise to a certain open string field theory,
then the open string field theory on the brane living
on the new background is given by changing
the start product
\be \lb{star}
f *_\gamma  g   \longrightarrow e^{ i\pi \gamma (Q^1_f Q^2_g - Q^2_f Q^1_g) } f *_0 g
\ee
where $*_0$ is the original star product and $Q^i_{f,g}$ are the $U(1)$ charges
of the fields $f$ and $g$. If one consider branes sitting at the origin,
the transformation (\ref{star}) does not lead to a
non-commutative field theory at low energies, because the $U(1)$ directions
are global symmetries of the field theory.
The effect of this transformation for the field theory living on a brane is
just to introduce certain phases in the lagrangian according to the rule in (\ref{star}).
If we know the gravity dual of the field theory living on a D-brane in the original background,
then the gravity dual of the deformed field theory corresponding to the D-brane on
the new background will be obtained by performing the $SL(2,R)$ transformation
on the original solution.

\section{Discussion}

   Along the present work, new and old examples of toric metrics
with holonomy $G_2$ has been presented. The direct sum of such
metrics with the flat four dimensional Minkowski one are the most
general solutions of the eleven dimensional supergravity which give
rise to IIA backgrounds satisfying the strong supersymmetry
condition (or Kahler condition) $F^{(1,1)}=0$ of reference
\cite{Kaste}. The equivalence between the formalisms of \cite{Kaste}
and the Apostolov-Salamon one \cite{Apostol} has been explained in
detail. In some sense, the statement that Apostolov-Salamon metrics
\cite{Apostol} "solve" the conditions of \cite{Kaste} (that is, the
holomorphic monopole equation and the strong supersymmetry
condition) could be a little misleading, because the general
solution of Apostolov-Salamon evolution equations is not known. We
just passed from one formalism into another, and presented some
simple examples. Nevertheless, the formalism of \cite{Apostol} has
the advantage that the presence of the toric isometry group is
immediately seen. Another interesting feature of this formalism is
that, as explained in section 3.2, \emph{any} 4-dimensional
hyperkahler metric can be extended to one with holonomy in $G_2$ by
means of a \emph{linear} system. Surprisingly, if the trivial flat
hyperkahler metric is used in this construction, the resulting
metric has irreducible curvature tensor and holonomy exactly $G_2$.
These examples are all related to half-flat six dimensional
structures, by means of Hitchin equations. An asymptotically
Calabi-Yau $G_2$ metric related to the flat hyperkahler metric has
also been presented, but the equations corresponding to other
hyperkahler basis is non linear in general and more difficult to
solve.

      Conditions in order to have $T^3$ instead of $T^2$ have been worked out.
One possible way to construct a $G_2$ holonomy metric with $T^3$
isometry is to use an hyperkahler basis with a
\emph{tri-holomorphic} isometry. These spaces correspond to
11-dimensional supergravity solutions for which the $\gamma$
deformation technique can be applied. This was done to our examples
and new supergravity solutions with four dimensional fluxes turned
on were found. The deformation technique was used as a solution
generating technique only, because we do not know the gravity duals
of our backgrounds.

     It is not clear for us whether or not it exist a coordinate system
for which these $G_2$ holonomy metrics are asymptotically conical.
Therefore we ignore if these metrics are suitable for obtaining
chiral matter after compactification to four dimensions. But even if
this is not so, there are many potential applications. To analyze
supersymmetry breaking mechanisms by the presence of non zero flux
of M-theory compactified on our manifolds \cite{Beasley}, to find
the membrane dynamics on our manifolds, to find the conserved
quantities associated with these isometries and to investigate $N=1$
dual theories \cite{Bozhilov}-\cite{Nunezo} could be some of them.
Moreover, these $G_2$ manifolds give rise to a dual theory in $3+1$
dimensions with minimal supersymmetry and extra KK modes. The
techniques in \cite{Lunin} could be useful in order to determine
which of these modes are relevant and which are not \cite{Gursoy}.

      There exist also applications related to the construction of
a topological string theory in seven dimensions \cite{Deboer} and to
the study of domain wall solutions \cite{Lukas}-\cite{Mohaupt}. The
half-flat backgrounds presented here are of relevance to type IIB
and heterotic string compactifications \cite{Gurro}-\cite{Gurro2}.
Another interesting task could be to see if it is possible to lift
these metrics to $Spin(7)$ ones as by use of the methods of
\cite{Aalok}. Further applications can be found in
\cite{Rajpoot}-\cite{Papadopolus}. We will return to some of these
points in a future investigation.
\\

   I sincerely acknowledge to V.Apostolov for many valuable
explanations.
\\

\textbf{Note added} When this work was finished, there appeared the
references \cite{Waldo}-\cite{Bozho}. Perhaps the results presented
in along our work have applications related to these references, and
also to reference \cite{Aalok2}.

\end{document}